\documentclass[twocolumn]{aastex63}
\usepackage{mathtools}
\usepackage{natbib}
\setcitestyle{authoryear,open={(},close={)}}
\usepackage{txfonts}
\usepackage{lipsum, babel}
\usepackage[T1]{fontenc}
\usepackage{ae,aecompl}
\usepackage{tikz}
\usepackage{graphicx}
\usepackage{subfigure}
\usepackage{float}
\usepackage{amsmath}
\usepackage{amssymb}
\usepackage{cleveref}
\usepackage{physics}
\usepackage{empheq}
\usepackage{booktabs}
\usepackage{array}
\newcolumntype{R}[1]{>{\raggedleft\arraybackslash}p{#1}}
\newcolumntype{L}[1]{>{\raggedright\arraybackslash}p{#1}}
\usepackage{multirow}
\usepackage[flushleft]{threeparttable}
\usepackage{mathrsfs}
\usepackage{soul}
\newcommand{\MyDiamond}[1][fill=black]{
\begin{tikzpicture}[x=1.2ex,y=1.2ex,line width=.1ex,line join=round, yshift=0.0ex] \draw  [#1]  (0,.5) -- (.5,1) -- (1,.5) -- (.5,0);
\end{tikzpicture}
}

\newcommand{\MyPlus}[1][fill=black]
{
\begin{tikzpicture}[x=1.5ex,y=1.5ex,line width=0.5ex]
\draw[#1] (0,.5) -- (1,.5);
\draw[#1] (.5,0) -- (.5,1);
\end{tikzpicture}
}

\newcommand{\MyCross}[1][fill=black]
{
\begin{tikzpicture}[x=1.ex,y=1.ex,line width=0.5ex]
\draw[#1] (0,0) -- (1,1);
\draw[#1] (0,1) -- (1,0);
\end{tikzpicture}
}

\newcommand{\MySolidLine}[1][fill=black]
{
\begin{tikzpicture}[x=2.ex,y=1.ex,line width=0.5ex]
\draw[#1] (0,.5) -- (2.5,.5);
\end{tikzpicture}
}

\newcommand{\MyDashedLine}[1][fill=black]
{
\begin{tikzpicture}[x=2.ex,y=1.ex,line width=0.5ex]
\draw [#1,dashed] (0,.5) -- (2.5,.5);
\end{tikzpicture}
}

\newcommand{\MyDashedDottedLine}[1][fill=black]
{
\begin{tikzpicture}[x=2.ex,y=1.ex,line width=0.5ex]
\draw [#1,dash dot] (0,.5) -- (2.5,.5);
\end{tikzpicture}
}

\newcommand{\MyDottedLine}[1][fill=black]
{
\begin{tikzpicture}[x=2.ex,y=1.ex,line width=0.5ex]
\draw [#1,dotted] (0,.5) -- (2.5,.5);
\end{tikzpicture}
}

\begin{document}
\title{An adaptive mesh, GPU-accelerated, and error minimized special relativistic hydrodynamics code}

\author[0000-0002-1868-0660]{Po-Hsun Tseng}
\affiliation{Institute of Astrophysics, National Taiwan University, Taipei 10617, Taiwan}

\author[0000-0002-1249-279X]{Hsi-Yu Schive}
\affiliation{Institute of Astrophysics, National Taiwan University, Taipei 10617, Taiwan}
\affiliation{Department of Physics, National Taiwan University, Taipei 10617, Taiwan}
\affiliation{Center for Theoretical Physics, National Taiwan University, Taipei 10617, Taiwan}
\affiliation{Physics Division, National Center for Theoretical Sciences, Hsinchu 30013, Taiwan}

\author[0000-0003-2654-8763]{Tzihong Chiueh}
\affiliation{Institute of Astrophysics, National Taiwan University, Taipei 10617, Taiwan}
\affiliation{Department of Physics, National Taiwan University, Taipei 10617, Taiwan}
\affiliation{Center for Theoretical Physics, National Taiwan University, Taipei 10617, Taiwan}

\correspondingauthor{Po-Hsun Tseng}
\email{zengbs@gmail.com}


\begin{abstract}
We present a new special relativistic hydrodynamics (SRHD) code capable of handling coexisting ultra-relativistically hot and non-relativistically cold gases. We achieve this by designing a new algorithm for conversion between primitive and conserved variables in the SRHD solver, which incorporates a realistic ideal-gas equation of state covering both the relativistic and non-relativistic regimes. The code can handle problems involving a Lorentz factor as high as $10^6$ and optimally avoid the catastrophic cancellation. In addition, we have integrated this new SRHD solver into the code \textsc{gamer} (\url{https://github.com/gamer-project/gamer}) to support adaptive mesh refinement and hybrid OpenMP/MPI/GPU parallelization. It achieves a peak performance of $7\times 10^{7}$ cell updates per second on a single Tesla P100 GPU and scales well to 2048 GPUs. We apply this code to two interesting astrophysical applications: (a) an asymmetric explosion source on the relativistic blast wave and (b) the flow acceleration and limb-brightening of relativistic jets.
\end{abstract}

\section{Introduction}
 Many high energy astrophysical problems involve relativistic flows. The problems include, for example, collimated jets in active galactic nuclei (AGN) (\citealt{Chiueh-1991ApJ...377..462C}; \citealt{Chiueh-1992ApJ...394..459L}; \citealt{Blandford2018}), collapsar models of long-duration gamma-ray bursts \citep{LongGRB}, magnetized relativistic winds and nebulae from pulsars (\citealt{1984ApJ...283..694K}; \citealt{1984ApJ...283..710K}; \citealt{Chiue-PhysRevLett.63.113}; \citealt{Chiueh_1998}), and mildly relativistic wide-angle outflows in neutron star mergers (\citealt{NM2}; \citealt{NM1}; \citealt{NM3}; \citealt{NM4}). The full scope of these problems generally involves substantial temperature changes between jets (winds) and ambient gases. For this reason, the pioneering works of \cite{Taub}, \cite{TM_EOS}, and \cite{Compare_TM_EOS} suggested Taub-Mathews equation of state (TM EoS) that approximates the exact EoS \citep{Synge} for ultra-relativistically hot (high-$T$ hereafter) gases coexisting with non-relativistically cold (low-$T$ hereafter) gases. 
 
 In addition, \cite{Noble_2006} first compared the accuracy of several schemes for recovering primitive variables in the Riemann problems by means of self-checking tests (see Appendix \ref{Appendix:Numerical error analysis} for details). \cite{NR_Limit} further proposed an inversion scheme for an arbitrary EoS and suggested that directly evolving the reduced energy density (i.e. the energy density subtracting the rest mass energy density from the total energy density) can avoid catastrophic cancellation in the non-relativistic limit. However, very few studies have systematically investigated how serious the catastrophic cancellation bears upon simulation results. This is partially due to the lack of exact solutions with which numerical results can be compared.
 
 In this paper, we propose a new numerical scheme for conversion between primitive and conserved variables in the presence of both high-$T$ and low-$T$ gases. The new scheme is carefully tailored to avoid catastrophic cancellation. To verify its accuracy, we numerically derive the exact solutions of two relativistic Riemann problems with the TM EoS and compare with the simulation results. It demonstrates that our new special relativistic hydrodynamics (SRHD) code can minimize numerical errors compared with conventional methods.
 
 We have integrated this new SRHD solver into the code \textsc{gamer} (\citealt{gamer-1}; \citealt{gamer-2}) to facilitate GPU acceleration and adaptive mesh refinement (AMR). This new code, \textsc{gamer-sr}, yields good weak and strong scalings using up to 2048 GPUs on \texttt{Piz-Daint}, the supercomputer at the Swiss National Supercomputing Centre (CSCS). Finally, we present two astrophysical applications, an asymmetric explosion and self-accelerating jets, to demonstrate the capability of this new code in extreme conditions. All simulation data are analysed and visualized using the package \texttt{yt} \citep{YT}.
 
 This paper is organized as follows. We introduce the equation of state and our new scheme for conversion between primitive and conserved variables in Section \ref{Relativistic Hydrodynamic Equations}. In Section \ref{Numerical Method}, we describe numerical methods, including the AMR structure, GPU acceleration, flexible time-steps, and correction of unphysical results. In Sections \ref{Test Problems} and \ref{Performance scaling}, we conduct numerical experiments to demonstrate the accuracy in both the non-relativistic (NR) and ultra-relativistic (UR) limits, the performance scalability, as well as the limitation of \textsc{gamer-sr}. Finally, we present two astrophysical applications in Section \ref{Astrophysical Applications} and draw the conclusion in Section \ref{conclusions}.
 
 Note that the speed of light and the Boltzmann constant are hard-coded to $1$ in \textsc{gamer-sr}. However, these physical constants are retained in this paper, except in Appendices, for dimensional consistency.

\section{Relativistic hydrodynamic Equations}
\label{Relativistic Hydrodynamic Equations}
\subsection{Relativistic hydrodynamics}
\label{Relativistic Hydrodynamics}
Mass and energy-momentum conservation laws of a special relativistic ideal fluid follow
\begin{subequations}
\label{eq:conservation laws}
\begin{align}
&\partial_{\nu}\left(\rho U^{\nu}\right)=0, \label{eq:number conservation}\\
&\partial_{\nu}T^{\mu \nu} = 0, \label{eq:energy conservation}
\end{align}
\end{subequations}
where
\begin{equation}
T^{\mu \nu} = \rho h U^{\mu} U^{\nu}/c^2 + p \eta^{\mu \nu}.
\end{equation}
$\rho$ and $p$ are the proper mass density and the pressure, $U^\mu$ the four-velocity, $\eta^{\mu \nu}$ the metric tensor of Minkowski space, and $c$ the speed of light. $h$ is the specific enthalpy, related to the specific thermal energy $\epsilon$ by
\begin{equation}
  h=c^2+\epsilon +\frac{p}{\rho}.
  \label{eq:eos}
\end{equation}

An equation of state, $h\left(\rho, p\right)$, is required to close \Cref{eq:conservation laws} and will be discussed in Section \ref{EoS}.
Throughout this paper, lower-case Greek indices run from 0 to 3, Latin ones from 1 to 3, and the Einstein summation convention is used, except when stated otherwise.

\Cref{eq:conservation laws} can be rewritten into a convenient conservative form for numerical integration:
\begin{subequations}
  \label{conservative form}
  \begin{align}
   &\partial_{t} D+\partial_{j} \left(DU^{j}/\gamma\right)=0,\label{D evolution}\\
   &\partial_{t}M^{i}+\partial_{j} \left(M^{i}U^{j}/\gamma+p\delta^{ij}\right)=0,\label{M evolution}\\
   &\partial_t E+\partial_j  \left(M^{j}c^2\right)=0, \label{E evoltion}
  \end{align}
\end{subequations}
where $\gamma$ is the Lorentz factor, and $\delta^{ij}$ is the Kronecker delta notation.

The five conserved quantities $D$, $M^{i}$, and $E$ are the mass density, the momentum densities, and the total energy density, respectively. All conserved variables are related to primitive variables ($\rho, U^{i}, p$) through
\begin{subequations}
  \begin{align}
    &D=\rho\gamma,\label{density}\\ 
    &M^{i}=Dh U^{i} /c^2,\label{momentum}\\
    &E=D h\gamma-p. \label{definition of reduced energy}
  \end{align}
  \label{relation between prim and cons}
\end{subequations}
Nevertheless, \cite{NR_Limit} suggest evolving the reduced energy density,
\begin{equation}
\tilde{E} \coloneqq E-Dc^2, \label{ETilde}
\end{equation}
instead of the total energy density; otherwise, extraction of a tiny thermal energy for a cold gas from the total energy will lead to catastrophic cancellation. An intuitive approach is to subtract \Cref{D evolution} from \Cref{E evoltion} so that we can obtain a new energy equation. However, the new energy flux, $\left(M^{j}-DU^{j}/\gamma\right)c^2$, also suffers from catastrophic cancellation in the NR limit. An appropriate new energy flux avoiding such a problem is $(\tilde{E}+p)U^{j}/\gamma$, which is mathematically equivalent to $\left(M^{j}-DU^{j}/\gamma\right)c^2$. The reduced energy equation for numerical integration can thus be cast into
\begin{equation}
    \partial_t \tilde{E}+\partial_j \left[\left(\tilde{E}+p\right)U^{j}/\gamma\right]=0,
    \label{ETilde evolution}
\end{equation}
which is to replace \Cref{E evoltion}.


Moreover, solving the Lorentz factor $\gamma$ as three-velocity ($v=\sqrt{v^{i}v_{i}}$) approaches $c$ can seriously suffer from catastrophic cancellation when using $\gamma=1/\sqrt{1-v^{i}v_{i}/c^2}$. Therefore, we explicitly adopt four-velocities ($U^{i}$) instead of three-velocity ($v^{i}$) for numerical computations and solve the Lorentz factor in terms of four-velocities as
\begin{equation}
\label{eq:new expression of Lorentz factor}
\gamma=\sqrt{1+U^iU_i/c^2},
\end{equation}
by which significant digits in $\gamma$ can be kept when $\gamma \gg 1$. 

In addition, unlike the three-velocity bounded by $c$, four-velocity $U^i$ has no upper limit and therefore can greatly reduce the risk of having $v>c$ due to numerical errors.

\subsection{Equations of state}
\label{EoS}
\textsc{gamer-sr} supports two kinds of EoS, the Taub-Mathews EoS (TM; \citealt{Taub}, \citealt{TM_EOS}, \citealt{Compare_TM_EOS}) and the polytropic EoS with a constant ratio of specific heats $\Gamma$. Assuming an ideal fluid in local thermal equilibrium and obeying the non-degenerate Maxwell-J\"{u}ttner statistics \citep{Juttner}, the exact EoS \citep{Synge} derived from the kinetic theory of relativistic gases is given by
\begin{equation}
    \frac{h_{\text{exact}}}{c^2}=
    \frac{K_{3}\left(mc^2/k_{B}T\right)}
         {K_{2}\left(mc^2/k_{B}T\right)}, 
\label{EXACT_EOS}
\end{equation}
where $k_{B}$ and $T$ are the Boltzmann constant and temperature, respectively, and $K_{n}$ the $n$-th order modified Bessel function of the second kind. However, direct use of \Cref{EXACT_EOS} is computationally inefficient because the evaluation of Bessel function is numerically expensive.

Alternatively, the TM EoS is an approximation of \Cref{EXACT_EOS} and given by
\begin{equation}
\frac{h_{\text{TM}}}{c^2}=2.5\left(\frac{k_B T}{mc^2}\right)+\sqrt{2.25{\left(\frac{k_B T}{mc^2}\right)}^{2}+1}.
\label{TM EOS}
\end{equation}

\begin{figure}
	\includegraphics[width=\columnwidth]{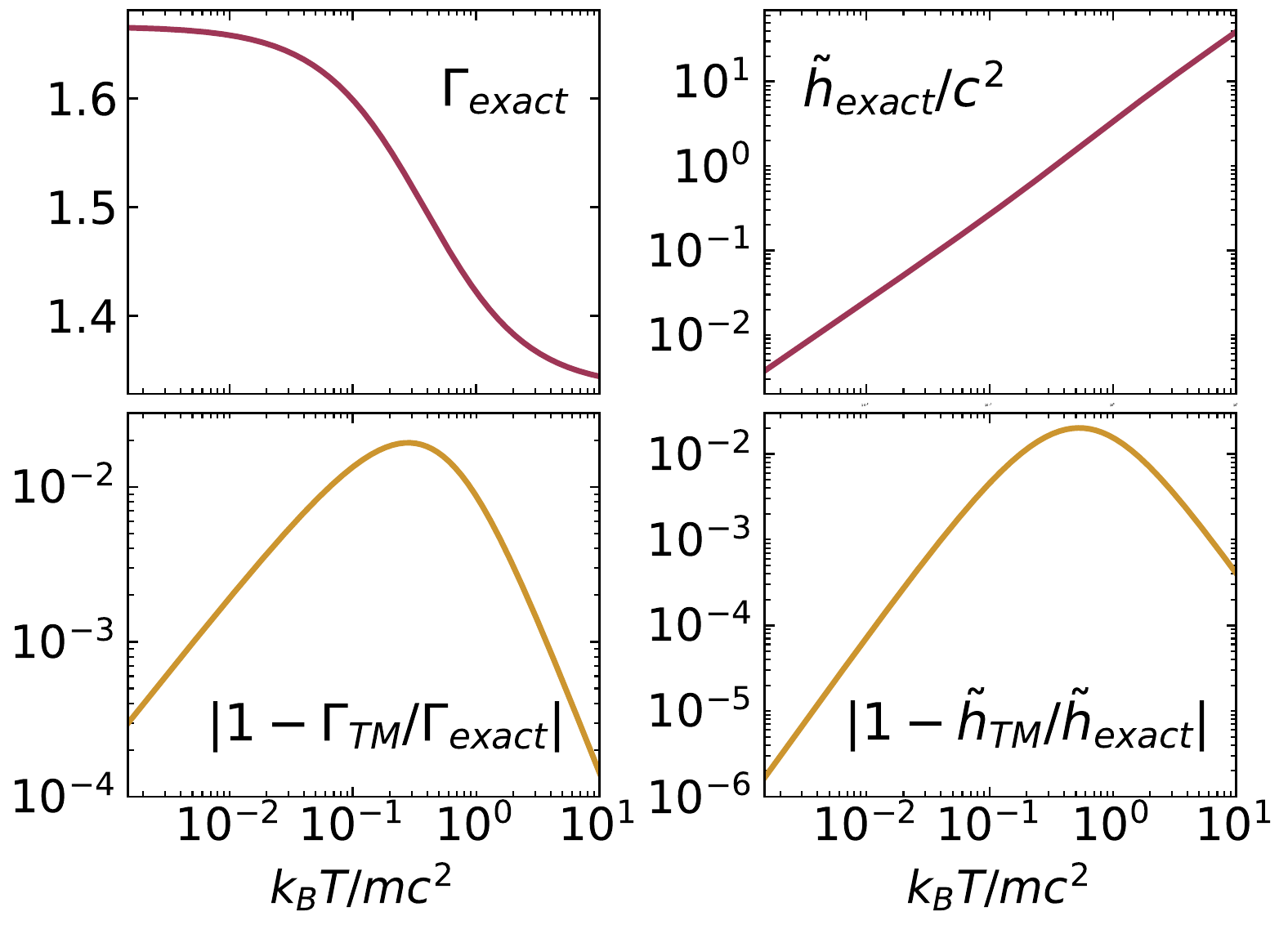}
    \caption{The effective adiabatic index $\Gamma$ (top left), the reduced enthalpy $\tilde{h}/c^2\coloneqq h/c^2-1$ (top right) as a function of temperature. Bottom panels show that \Cref{TM EOS} approaches \Cref{EXACT_EOS} in both high- and low-$T$ limits, where the maximum relative errors $\abs{1-\Gamma_{\text{TM}}/\Gamma_{\text{exact}}}$ and $\abs{1-\tilde{h}_{\text{TM}}/\tilde{h}_{\text{exact}}}$, are only 1.9 and 2.0 per cent, respectively.}
   \label{fig:compare_eos}
\end{figure}

The effective $\Gamma$ can be found by equating \Cref{EXACT_EOS} or \Cref{TM EOS} to the polytropic EoS,
\begin{equation}
    \frac{h_{\Gamma}}{c^2}=
    1+\frac{\Gamma}{\Gamma-1}\left(\frac{k_{B}T}{mc^2}\right),
\end{equation}
and solving $\Gamma$ for the exact or TM EoS, respectively. As depicted in \Cref{fig:compare_eos}, the maximum relative errors $\abs{1-\Gamma_{\text{TM}}/\Gamma_{\text{exact}}}$ and $\abs{1-\tilde{h}_{\text{TM}}/\tilde{h}_{\text{exact}}}$ are found to be only 1.9 and 2.0 per cent, respectively.  In addition, \Cref{TM EOS} approaches \Cref{EXACT_EOS} in both high- and low-$T$ limits. Detailed comparisons between \Cref{EXACT_EOS} and \Cref{TM EOS} have been presented previously (\citealt{Compare_TM_EOS}; \citealt{RC_EOS}; \citealt{NR_Limit}) and we do not repeat here. 

On the other hand, the polytropic EoS has the advantage of simplicity and therefore has been used in many SRHD codes, such as \textsc{flash} \citep{FLASH}, \textsc{cafe} \citep{CAFE}, and \textsc{xtroem-fv} \citep{XTROEM}. However, the polytropic EoS cannot handle the case where relativistic gases  and non-relativistic gases coexist, primarily because the ratio of specific heats depends sensitively on temperature when $k_{B}T \sim mc^2$ (see the upper left panel in \Cref{fig:compare_eos}). Moreover, the polytropic EoS with a non-relativistic $\Gamma=5/3$ and a relativistic $\Gamma=4/3$ does not satisfy the Taub's fundamental inequality for ideal gases \citep{Taub}
\begin{equation}
    \left[\frac{h}{c^2}-\left(\frac{k_B T}{mc^2}\right)\right]
    \left[\frac{h}{c^2}-4\left(\frac{k_B T}{mc^2}\right)\right]
    \geq 1,
\end{equation}
implying that $\Gamma$ must lie between $4/3$ and $5/3$ for any positive and finite value of temperature. Although the polytropic EoS is physically incorrect, we still reserve this feature in \textsc{gamer-sr} for fast computation of a pure non-relativistic or relativistic gas.

The other two important quantities are the Mach number ($\mathscr{M}$) and the sound speed ($c_{s}$), given by
\begin{equation}
    \mathscr{M}=\frac{\sqrt{U^iU_i}}{U_{s}},
    \label{eq:MachNumber}
\end{equation}
and
\begin{equation}
\frac{c_{s}}{c} = \sqrt{\frac{k_{B}T/mc^{2}}{3h/c^2}\left(\frac{5h/c^2-8k_{B}T/mc^{2}}{h/c^2-k_{B}T/mc^{2}}\right)},
\label{sound_speed}
\end{equation}
for the TM EoS, where $U_{s}=c_{s}/\sqrt{1-\left(c_{s}/c\right)^2}$. The sound speed approaches $c/\sqrt{3}$ at ultra-relativistic temperature and will be used in the Riemann solver.

\subsection{Conversion between primitive and conserved variables}
\label{section: Conversion between primitive variables and conserved ones}
In standard Riemann-type numerical schemes, conversion between conserved and primitive variables is a common procedure for data reconstructions and flux computations. For non-relativistic hydrodynamics, this conversion can be carried out in a straightforward and analytical manner. However, designing an accurate and efficient conversion algorithm for a relativistic problem in the presence of NR gases, which involves root-finding, is challenging. This is because catastrophic cancellations may arise in the non-relativistic gas. 

Here we propose a new conversion scheme to solve this problem based on the TM EoS. The reduced energy density (Equation \ref{ETilde}) and the momentum density (Equation \ref{momentum}) satisfy the relation
\begin{equation}
\begin{aligned}
&\left(\frac{\tilde{E}}{Dc^2}\right)^2+2\left(\frac{\tilde{E}}{Dc^2}\right)-\left(\frac{\abs{\mathbf{M}}}{Dc}\right)^2
\\&= \frac{\tilde{h}^2}{c^4}+\frac{2\tilde{h}}{c^2}-2\left(\frac{k_{B}T}{mc^2}\right)\left(\frac{\tilde{h}}{c^2}+1\right)+\frac{\left(k_{B}T/mc^2\right)^2\left(\tilde{h}+c^2\right)^2}{\left(\tilde{h}+c^2\right)^2+\left(\frac{\abs{\mathbf{M}}c}{D}\right)^2}\\
&\coloneqq f\left(\tilde{h}\right),
\end{aligned}
\label{transcendental equation}
\end{equation}
where $f$ is positive definite, $\tilde{h}\coloneqq h-c^2$ is the reduced enthalpy, and the temperature $k_{B}T/mc^2$ is related to $\tilde{h}$ by inverting \Cref{TM EOS}:
\begin{equation}
\frac{k_{B}T}{mc^2}=\frac{2\left(\tilde{h}/c^2\right)^2+4\left(\tilde{h}/c^2\right)}{5\left(\tilde{h}/c^2\right)+5+\sqrt{9\left(\tilde{h}/c^2\right)^2+18\left(\tilde{h}/c^2\right)+25}}.
\label{T of h}
\end{equation}

The conserved variables ${\tilde E}$, $M^{j}$, and $D$ on the left-hand side are known quantities updated at every time step, from which one can solve for $\tilde{h}$.

We adopt $\tilde{h}=h-c^2$ instead of $h$ as the root because the latter is dominated by rest mass energy density in the low-$T$ limit and thus will suffer from catastrophic cancellation when numerically extracting temperature from trailing digits. 

Equation~(\ref{transcendental equation}) is suitable for the Newton-Raphson iteration method as it is a monotonically increasing function of $\tilde{h}$. That is, \Cref{transcendental equation} has no zero derivative of $\tilde{h}$ that might otherwise lead to a divergence of the iterative procedure. The Newton-Raphson method requires an initial guess of $\tilde{h}$ and the derivative of \Cref{transcendental equation} for iteration, both of which are presented in Appendix \ref{The choice of initial guesses}.

After obtaining $\tilde{h}$, we substitute it into Equation~(\ref{momentum}) to get four-velocity:
\begin{equation}
U^i=\frac{M^{i}c^2}{D\left(c^2+\tilde{h}\right)}.\label{eq:four-velocity}
\end{equation}
Next, we compute the Lorentz factor and proper mass density from Equation~(\ref{eq:new expression of Lorentz factor}) and Equation~(\ref{density}) and then use Equation~(\ref{T of h}) to obtain temperature. Finally, the pressure is given by
\begin{equation}
    p=\rho c^2 \left(\frac{k_{B}T}{mc^2}\right).
    \label{eq:p=rhoT}
\end{equation}

Justifying the superiority of our new conversion scheme using $\tilde{E}$, we estimate the relative error of computing $a-b$ by \citep{Nicholas}
\begin{equation}
\frac{\left|a\right|+\left|b\right|}{\left|a-b\right|}\epsilon_{\text{machine}}, \label{error bound of subtraction}
\end{equation}
where $\epsilon_{\text{machine}}$ is the machine round-off error. Thus, the error of the new conversion scheme can be estimated by  substituting $\left[\left(\tilde{E}/Dc^2\right)^2+2\left(\tilde{E}/Dc^2\right)\right]$ and $\left(\abs{\mathbf{M}}/Dc\right)^2$ for $a$ and $b$, respectively, in \Cref{error bound of subtraction}. The error in terms of primitive variables reads
\begin{equation}
\begin{aligned}
\label{eq:ImprovedRelativeError}
&\left[\frac{\gamma^2\left(\tilde{h}+1\right)^2\left(1+\beta^2\right)+\frac{T^2}{\gamma^2}-2\left(\tilde{h}+1\right)T-1}{\left(\tilde{h}+1\right)^2+\frac{T^2}{\gamma^2}-2\left(\tilde{h}+1\right)T-1}\right]\epsilon_{\text{machine}}\\
&\approx \left(1+\mathscr{M}^2\right)\epsilon_{\text{machine}}.
\end{aligned}
\end{equation}
where $\beta=\sqrt{v^{i}v_{i}}/c$.
The approximate equality in \Cref{eq:ImprovedRelativeError} holds for all finite temperature.

For the original scheme using the total energy density $E$ instead of $\tilde{E}$, a similar error estimation can be performed by replacing $\tilde{E}$ with $E-Dc^2$ on the left-hand side of \Cref{transcendental equation}, which gives
\begin{equation}
    \left[\frac{2\gamma^2\left(\tilde{h}+1\right)^2+\frac{T^2}{\gamma^2}-2\left(\tilde{h}+1\right)T+\left(\tilde{h}+1\right)^2+1}{\left(\tilde{h}+2\right)\tilde{h}+\frac{T^2}{\gamma^2}-2\left(\tilde{h}+1\right)T}\right]\epsilon_{\text{machine}}.
\label{eq:OriRelativeError}
\end{equation}

\begin{figure}
	\includegraphics[width=\columnwidth]{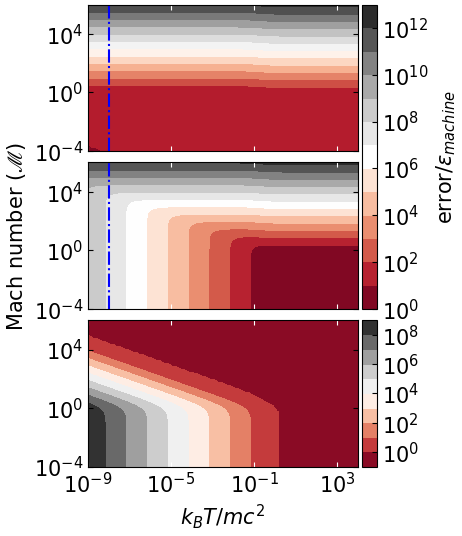}
    \caption{Numerical errors of the conversion from conserved to primitive variables as a function of $\mathscr{M}$ and $k_{B}T/mc^2$. The top and middle panel show the errors of the new and original schemes estimated by \Cref{eq:ImprovedRelativeError} and \Cref{eq:OriRelativeError}, respectively. The bottom panel shows the ratio of \Cref{eq:OriRelativeError} to \Cref{eq:ImprovedRelativeError}. \Cref{fig:ErrorAnalysis} in Appendix \ref{Appendix:Numerical error analysis} provides numerical evidences showing a remarkable consistency with the predicted values at $k_{B}T/mc^2=10^{-8}$ (blue dashed-dotted line).}
   \label{fig:ErrorDistribution}
\end{figure}

\Cref{fig:ErrorDistribution} shows the contour plots of \Cref{eq:ImprovedRelativeError} for the new scheme (top panel) and \Cref{eq:OriRelativeError} for the original scheme (middle panel) as a function of $\mathscr{M}$ and temperature. The bottom panel shows the ratio of \Cref{eq:OriRelativeError} to \Cref{eq:ImprovedRelativeError}. It demonstrates the advantage of using $\tilde{E}$. The top panel shows that using $\tilde{E}$ in the conversion scheme is almost error-free when dealing with subsonic flows at any finite temperature, including the low-$T$ limit. In supersonic flows, the numerical errors proportional to $\mathscr{M}^2$ are common and caused by finite digits of floating numbers. In comparison, the middle panel shows the error using $E$, which severely suffers from catastrophic cancellation in the low-$T$ limit even when $\mathscr{M}\ll 1$. See also \Cref{fig:ErrorAnalysis} in Appendix \ref{Appendix:Numerical error analysis}.

On the other hand, conversion from primitive to conserved variables is also needed in the Riemann solver. This procedure involves straightforward substitution without the need of root-finding. We use
\begin{equation}
\frac{\tilde{h}}{c^2} = 2.5\left(\frac{k_{B}T}{mc^2}\right)+\frac{2.25\left(k_{B}T/mc^2\right)^2}{1+\sqrt{2.25\left(k_{B}T/mc^2\right)^2+1}},
\label{eq:T_to_HTilde}
\end{equation}
and
\begin{equation}
\frac{\tilde{E}}{Dc^2} = \frac{\left(\frac{\abs{\mathbf{M}}}{Dc}\right)^2+f\left(\tilde{h}\right)}{1+\sqrt{1+\left(\frac{\abs{\mathbf{M}}}{Dc}\right)^2+f(\tilde{h})}},
\label{eq:E_D}
\end{equation}
to compute $\tilde{h}$ and $\tilde{E}$, where $f(\tilde{h})$ can be computed from \Cref{transcendental equation} with known $\abs{\mathbf{M}}/Dc$. Note that \Cref{eq:T_to_HTilde} and \Cref{eq:E_D}, following directly from \Cref{TM EOS} and \Cref{transcendental equation} without any approximation, are written in a form without any subtraction to avoid catastrophic cancellation. In contrast, using \Cref{definition of reduced energy} and \Cref{ETilde} to compute the reduced energy density $\tilde{E}$ can suffer from catastrophic cancellation in the NR limit.

We close this section by providing a flowchart of the new conversion scheme in \Cref{fig:flowchart} in Appendix \ref{Appendix:Numerical error analysis} and by summarizing the equations actually solved by \textsc{gamer-sr}. Other mathematically equivalent forms are unrecommended as they may suffer from catastrophic cancellation in the UR or NR limit.
\begin{itemize}
\item Evolution equations: Equation~(\ref{D evolution}, \ref{M evolution}, \ref{ETilde evolution}).
\item Lorentz factor:
\Cref{eq:new expression of Lorentz factor}.
\item Four-velocities: \Cref{eq:four-velocity}.
\item Temperature: \Cref{T of h}. 
\item Pressure: \Cref{eq:p=rhoT}.
\item Reduced enthalpy: \Cref{eq:T_to_HTilde}.
\item Reduced energy density: \Cref{eq:E_D}.
\end{itemize}

\section{Numerical Methods}
\label{Numerical Method}
\subsection{A GAMER Primer}
Due to the flexibility and extensibility of \textsc{gamer} (\citealt{gamer-1}; \citealt{gamer-2}), the SRHD module directly inherits the AMR structure and the MPI/OpenMP/GPU parallelization framework of hydrodynamics, and therefore we only provide a summary here. We define the base grid resolution as level-0 and the $\ell$th refinement as level-$\ell$, where level-$\ell$ has a spatial resolution $2^{\ell}$ times higher than that of the base level. Data in \textsc{gamer} are always decomposed into patches, each of which consists of $8^3$ cells, and the AMR implementation is realized by constructing a hierarchy of patches in an octree structure. According to user-defined refinement criteria, we can create or remove fine patches under the proper-nesting constraint. 

In addition to the refinement criteria provided by the hydrodynamics module, we also implement two refinement criteria for SRHD: the gradient of the Lorentz factor and the magnitude of $|\mathbf{M}|/D$. The former aims to capture the thin and high-$\gamma$ shell in the Sedov-Taylor blast wave, while the latter ensures that the spine region in an over-pressured jet (cf. \Cref{fig:Limb_brightened_jet}) can be fully resolved. For all refinement criteria, the refinement thresholds on different levels can be set independently as run-time parameters.

We port the routines involving massive floating-point operations to GPUs such as the SRHD solvers and time-step calculations. On the other hand, we use CPUs to perform ghost-zone interpolation and patch refinement. As a result, we recommend using the refinement criteria only involving conserved variable for better performance because conserved variables are readily available from memory. By contrast, primitive variables can only be obtained by root-finding iteration, which is computationally expensive.

For enhancing software portability and reusability, GAMER not only supports both CPU-only and GPU modes but also allows the same physics modules to be shared by both CPU and GPU computations. Specifically, in the CPU-only mode, we compute different grid patches in the same MPI process in parallel with OpenMP. In the GPU mode, we replace these OpenMP parallel clauses with CUDA thread blocks and then use threads within the same thread block to update all cells within the same grid patch. This scheme maximizes the reuse of physics routines, avoids redundant code development and maintenance, and significantly lowers the barrier of code extension, especially for developers not acquainted with GPU programming. We have utilized this CPU/GPU integration infrastructure in the SRHD implementation.

\textsc{gamer-sr} supports the MUSCL-Hancock \citep{Toro} and VL (\citealt{VL1}; \citealt{VL2}) schemes for numerical integrations and a piece-wise linear method (PLM; \citealt{van_Leer_1979}) for data reconstruction. For the Riemann solver, it supports both relativistic HLLC and HLLE solvers (\citealt{HLLC_srhydro}; \citealt{HLLC_srmhd}), which have been adapted not only to be compatible with the TM EoS by using the corresponding sound speed, \Cref{sound_speed}, but also to evolve the reduced energy density (i.e. replacing $E$ with $\tilde{E}+Dc^2$).

\subsection{Flexible Time-step}
\textsc{gamer-sr} provides two Courant-Friedrichs-Lewy (CFL) conditions for time-step determination. The first one is based on the local signal propagation speed, $S_{\text{max}}$, which gives maximum allowed time-steps in a wide dynamical range. Thus, it can significantly improve performance when the maximum $v/c$ is not close to unity. The other is based on the speed of light, where we simply replace $S_{\text{max}}$ by $c$. It gives the most conservative estimation of time-steps and is more time-consuming when the flow speed is far less than $c$, although it is simple to implement and requires less computation.

To calculate $S_{\text{max}}$, we first define $\mathbf{\hat{u}_s}$ to be a spatial unit vector in the direction of sound propagation, we then apply the Lorentz boost with velocity $-\pmb{\beta}$ to the four-velocity of sound speed $(\gamma_s, U_s\mathbf{\hat{u}_s})$ from local rest frame to laboratory frame. We finally obtain the four-velocity of signal that travels in laboratory frame as follows:
\begin{equation}
    \left(\gamma\gamma_s+\gamma U_s\left(\pmb{\beta}\cdot\mathbf{\hat{u}_s}\right),U_{s}\mathbf{\mathbf{\hat{u}_s}}+\left(\gamma-1\right)U_s\left(\pmb{\hat{\beta}}\cdot\mathbf{\hat{u}_s}\right)\pmb{\hat{\beta}}+\pmb{\beta}\gamma\gamma_s\right),
    \label{time_step_propagation of information}
\end{equation}
where $\gamma$ and $\gamma_s$ are the Lorentz factor of flow and of sound speed. $U_s$ is the four-velocity of sound speed defined by $c_s/\sqrt{1-c_s^2}$. Since the direction of the fastest signal propagation is in general parallel to flow velocity, we assume that both sound and flow propagate in the same direction (i.e. $\mathbf{\hat{u}_s}=\pmb{\hat{\beta}}$). The spatial components of \Cref{time_step_propagation of information} then reduce to
\begin{equation}
    \left(\beta \gamma \gamma_s+\gamma U_s\right)\pmb{\hat{\beta}}.
    \label{time_step_x}
\end{equation}
Motivated by \Cref{time_step_x}, we simply choose $\abs{U^{i}}\gamma_s+\gamma U_s$ as the bound of each spatial component and sum over $\abs{U^{i}}\gamma_s+\gamma U_s$ for each spatial component to obtain
\begin{equation}
U_{\text{max}}
=\gamma_{s}\left(\abs{U_{x}}+\abs{U_{z}}+\abs{U_{z}}\right)+3\gamma U_s,
\label{time_step_U_max}
\end{equation}
where $U_{x/y/z}$ is the $x/y/z$-component of the four-velocities of flow.

Note that \Cref{time_step_U_max} is essentially the addition of flow speed and sound speed in special relativity theory. Converting \Cref{time_step_U_max} back to three-velocity

\begin{equation}
S_{\text{max}}=U_{\text{max}}/\sqrt{1+\left(U_{\text{max}}/c\right)^2}, \label{SMax}
\end{equation}

and substituting \Cref{SMax} into the CFL condition, we finally obtain the flexible time-step based on the local signal propagation speed for SRHD:
\begin{equation}
\label{time_step_sound_speed}
\Delta t = C_{\text{CFL}}\left(\frac{\Delta h}{S_{\text{max}}}\right),
\end{equation}
where $\Delta h$ is the cell spacing and $C_{\text{CFL}}$ the safety factor with a typical value of $\sim 0.5$ for MUSCL-Hancock and VL schemes.

Note that \Cref{time_step_sound_speed} can be reduced to its non-relativistic counterpart,
\begin{equation}
    \Delta t=C_{\text{CFL}}\left(\frac{\Delta h}{\abs{v_x}+\abs{v_y}+\abs{v_z}+3c_s}\right),
\end{equation}
when $\gamma\sim 1$ and to $\Delta t = C_{\text{CFL}}\Delta h/c$ when $\gamma \gg 1$.

\subsection{Handling unphysical results}
Unphysical results, for example, negative pressure, negative density and superluminal motion, can stem from the failure of the following criterion:
\begin{equation}
\left(\frac{\tilde{E}}{Dc^2}\right)^2+2\left(\frac{\tilde{E}}{Dc^2}\right)-\left(\frac{\abs{\mathbf{M}}}{Dc}\right)^2>\epsilon_{\text{machine}},
\label{Unphysical result}
\end{equation}
where the left-hand side involves the numerically updated quantities and $\epsilon_{\text{machine}}$ is the machine epsilon -- typically, $2\times10^{-16}$ for double precision and $1\times 10^{-7}$ for single precision. The failure may take place in one of the following four steps:

\emph{(1) SRHD solver}\\
SRHD solver is responsible for updating the conserved variables by a given time-step. If unphysical result occurs in a cell, we redo data reconstruction by reducing the original minmod coefficient by a factor of 0.75. If the failure still occurs, we further reduce the minmod coefficient repeatedly until \Cref{Unphysical result} passes or the reduced minmod coefficient vanishes. Note that interpolating with a vanished minmod coefficient is essentially equivalent to the piece-wise constant spatial reconstruction.

\emph{(2) Grid refinement}\\
Unphysical results may occur during grid refinement when performing interpolations on parent patches. The remedy here is the same as that in the SRHD solver. We repeat the interpolation process with a reduced minmod coefficient on the conserved variables until \Cref{Unphysical result} passes or the minmod coefficient vanishes. A vanished minmod coefficient is essentially equivalent to directly copying data from the parent patch without interpolation.

\emph{(3) Ghost-zone interpolation}\\
To preserve conservation, where the volume-weighted average of child patch data are equal to its parent patch data, we normally fill the ghost zones of the patches on level $\ell+1$ by interpolating the conserved variables on level $\ell$ when the ghost zones lie on level $\ell$. However, if unphysical results occur, we interpolate primitive variables instead. Interpolating primitive variables is more robust than interpolating conserved variables since \Cref{Unphysical result} is always satisfied. After interpolation, we fill the ghost zones with the conserved variables derived from the interpolated primitive variables. Note that this procedure still preserves conservation because ghost zones do not affect conservation.

\emph{(4) Flux correction operation.}\\
For a leaf coarse patch adjacent to a coarse-fine interface, the flux difference between the coarse and fine patches on the interface will be used to correct the coarse-patch conserved variables adjacent to this interface. If unphysical results are found after this flux correction, we simply ignore the correction on the failed cells. Skipping the correction will break the strict conservation but it only occurs rarely.

\section{Test Problems}
\label{Test Problems}
To understand how evolving the total energy density may deteriorate simulation results and to demonstrate how much the new scheme improves, we compare the results from evolving $E$ by the flux $M$ (original scheme) with that from evolving $\tilde{E}$ by the flux $(\tilde{E}+p)U_{x}/\gamma$ (new scheme). Since catastrophic cancellation is likely to occur in UR and NR limits, we will conduct several test problems in these two limits. All simulations throughout this paper adopt the HLLC Riemann solver and PLM data reconstruction unless otherwise specified.

\definecolor{NewHot}{RGB}{255, 216, 23}
\definecolor{NewCold}{RGB}{21, 143, 191}
\definecolor{OriHot}{RGB}{158,54,86}
\definecolor{OriCold}{RGB}{204, 149, 47}
\subsection{Convergence test for sinusoidal waves}
\label{section:convergence test}
We perturb proper mass density in the high- and low-$T$ limits to compare the accuracy of both schemes over a wide dynamical range. We construct the initial conditions as follows. All cases share homogeneous and static background with proper mass density $\rho_{0}=1$ on uniform grids, whereas the ambient temperatures are set to $k_{B}T/mc^2=10^{10}$ and $10^{-10}$ for the high- and low-$T$ limits, respectively. We then sinusoidally perturb the background with a tiny amplitude, $\delta \rho/\rho_{0}=10^{-6}$. 

To monitor how errors in the numerical solution decrease as a function of increasing spatial resolution in the three-dimensional space, we adopt a propagating wave along the diagonal direction of the simulation cubic box with the periodic boundary condition. Thus, the analytical solution is $\rho(\mathbf{x},t)=\rho_{0}+\delta\rho\sin \left[\left(x+y+z\right)/\sqrt{3}-c_{s}t\right]$, where $c_{s}$ is the sound speed given by Equation~(\ref{sound_speed}). 

\begin{figure}
\includegraphics[width=\columnwidth]{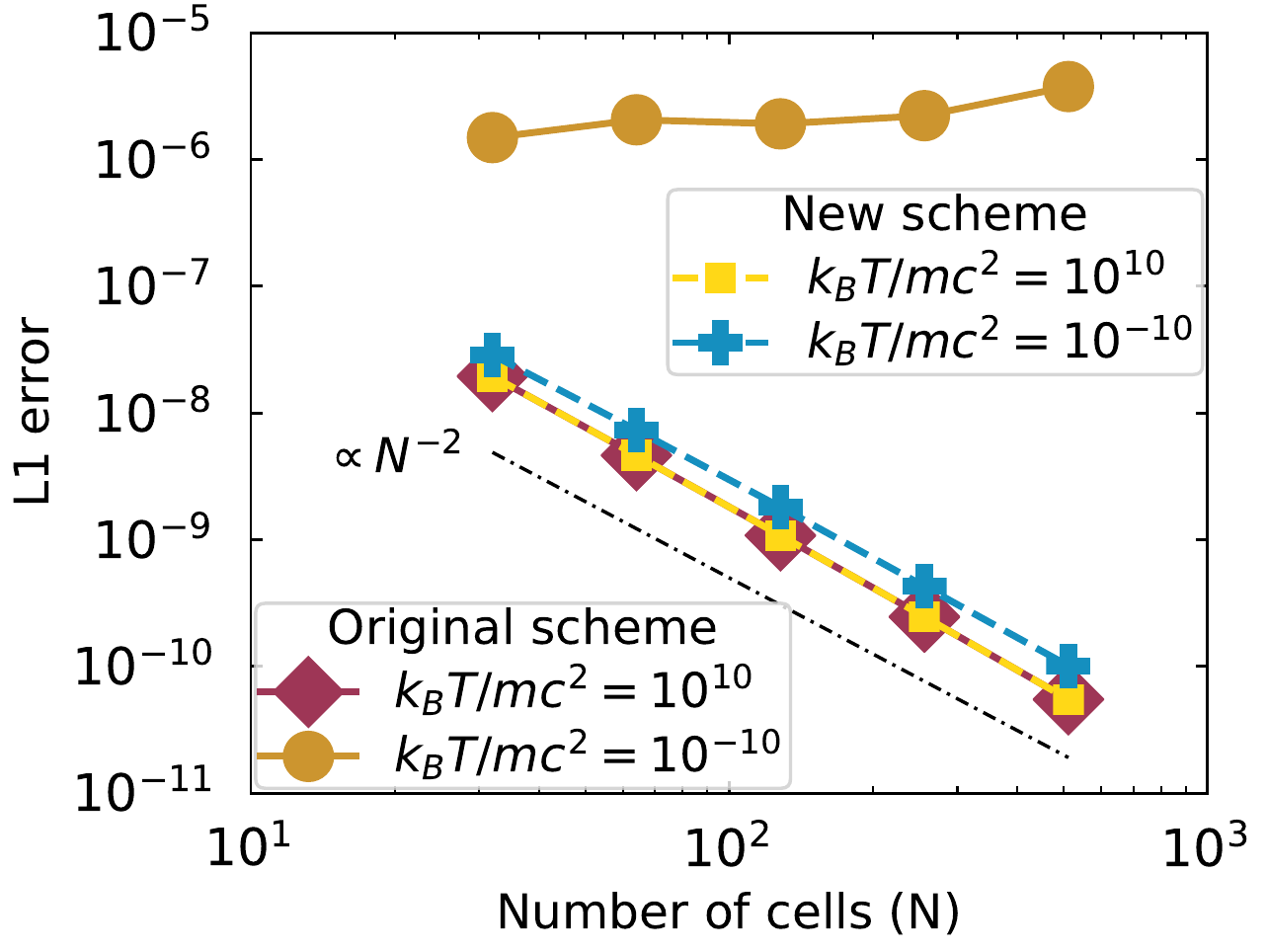}
\caption{Convergence test for sinusoidal waves. The L1 errors of the new scheme in both the high-$T$ limit ($\tikz\draw[black,fill=NewHot] (0,0) rectangle (0.15,0.15);$) and low-$T$ limit ($\MyPlus[draw=NewCold,fill=NewCold]$) decrease as $N^{-2}$, consistent with the second-order accuracy of the MUSCL-Hancock scheme with PLM data reconstruction. However, the L1 error of the original scheme in the low-$T$ limit (\tikz\draw[black,fill=OriCold] (0,0) circle (.5ex);) is much larger and roughly equal to a constant of $2\times 10^{-6}$. This is expected because the error in the original scheme can be estimated from \Cref{eq:error in the NR limit}. For the original scheme in the opposite high-$T$ limit ($\MyDiamond[draw=black,fill=OriHot]$), the discretization error, however, completely overwhelms the error ($\sim 4\times10^{-16}$) estimated from \Cref{eq:OriRelativeError} in the high-$T$ limit when $\gamma \sim 1$, thus dominating the L1 error.}
\label{fig:convergence test}
\end{figure}

We define the L1-norm error as
\begin{equation}
L1(Q)=\frac{1}{N}\sum^{N}_{i=1}\abs{1-\frac{Q_{\text{numerical}}\left(\mathbf{x_{i}}\right)}{Q_{\text{analytical}}\left(\mathbf{x_{i}}\right)}},
\label{relative L1 error}
\end{equation}
where $Q_{\text{numerical}}\left(\mathbf{x_{i}}\right)$ is the numerical solution of $i$-th cell at $\mathbf{x_{i}}$ and $Q_{\text{analytical}}\left(\mathbf{x_{i}}\right)$ is the corresponding analytical solution. 
We then calculate the L1 error of the proper mass density along the wave propagating direction.
As shown in \Cref{fig:convergence test}, the L1 errors of the new scheme in both the high-$T$ limit ($\tikz\draw[black,fill=NewHot] (0,0) rectangle (0.15,0.15);$) and low-$T$ limit ($\MyPlus[draw=NewCold,fill=NewCold]$) decrease as $N^{-2}$, consistent with the second-order accuracy of the MUSCL-Hancock scheme with PLM data reconstruction. However, the error of the original scheme in the low-$T$ limit (\tikz\draw[black,fill=OriCold] (0,0) circle (.5ex);) is much larger and roughly equal to a constant of $2\times 10^{-6}$. This is expected because the error arising from the original scheme can be estimated from \Cref{eq:OriRelativeError} in the NR limit:
\begin{equation}
    \frac{4}{3\left(\frac{k_{B}T}{mc^2}\right)}\epsilon_{\text{machine}},
    \label{eq:error in the NR limit}
\end{equation}
where $k_{B}T/mc^2=10^{-10}$ and $\epsilon_{\text{machine}}\sim 10^{-16}$ for double precision. 

We thus conclude that for the original scheme in the NR limit, the cancellation between $(E/Dc^2)^2$ and $[(\abs{\mathbf{M}}/Dc)^2+1]$ leads to an error of $\sim 10^{-6}$ when computing primitive variables, roughly consistent with the L1 error (\tikz\draw[black,fill=OriCold] (0,0) circle (.5ex);). For the opposite high-$T$ limit ($\MyDiamond[draw=black,fill=OriHot]$), the discretization error, however, completely overwhelms the error ($\sim 4\epsilon_{\text{machine}}\sim4\times 10^{-16}$) estimated from \Cref{eq:OriRelativeError} in the high-$T$ limit, thus dominating the L1 error. The error arising from the cancellation in the new scheme, $\left(\tilde{E}/Dc^2\right)^2+2\left(\tilde{E}/Dc^2\right)-\left(\abs{\mathbf{M}}/Dc\right)^2$, on the left side of \Cref{transcendental equation}, is close to $\epsilon_{\text{machine}}$ in both the high- and low-$T$ limits when $\mathscr{M}<1$ (see Section \ref{section: Conversion between primitive variables and conserved ones} and Appendix \ref{Appendix:Numerical error analysis} for details).

\subsection{1-D relativistic Riemann problems}
\label{subsec:1-DReletivisticRiemann problems}
The 1-D Riemann problem \citep{SOD1978} has played an important role by providing exact nonlinear solutions against which (relativistic) hydrodynamic codes can be tested. Riemann problem is an initial-value problem with a piece-wise constant initial data that has a single discontinuity in the domain of interest. In this subsection, we directly compare the new and original schemes by simulating two relativistic Riemann problems. We then demonstrate that the new scheme handles both the UR and NR limits very well. By contrast, the original scheme severely suffers from numerical errors in the NR limit. Both schemes share the same numerical setup, e.g., MUSCL-Hancock integration, PLM data reconstruction, hybrid van-Leer, generalized minmod slope limiter, and uniform grids with the outflow boundary condition. In addition, we have numerically derived the exact solution of a nontrivial relativistic Riemann problem with the TM EoS (see Appendix \ref{appendix:exact solution} for details) in order to verify the numerical results.

\begin{table*}[t]
\caption{The left and right initial states of the Riemann problems in Section \ref{subsec:1-DReletivisticRiemann problems}. We denote the left/right states by the subscript $L/R$.}
\label{tb:IC_RiemannProblems}
\begin{tabular}{@{}lrrrrrrr@{}}
\hline
                         & $p_{L}$   & $\rho_{L}$ & $U_{L}$   & $p_{R}$    & $\rho_{R}$ & $U_{R}$   & Floating-point format \\ \hline
Ultra-relativistic limit & $1.0$     & $10^{-5}$  & $10^6$    & $1.0$      & $10^{-5}$  & $-10^{6}$ & Double precision      \\ \hline
Mixed limits             & $10^{-4}$ & $10^{2}$   & $10^{-3}$ & $10^{-10}$ & $10^{-12}$ & $-10^{2}$  & Single precision      \\ \hline
\end{tabular}
\end{table*}

\definecolor{HeadOnCollisionExact}{RGB}{122, 24, 18}
\definecolor{HeadOnCollisionOri}{RGB}{179, 169, 62}
\definecolor{HeadOnCollisionNew}{RGB}{69, 177, 230}

\subsubsection{Ultra-relativistic limit}
\label{subsubsec:Ultra-relativistic limit}
\begin{figure*}
\includegraphics[width=\linewidth]{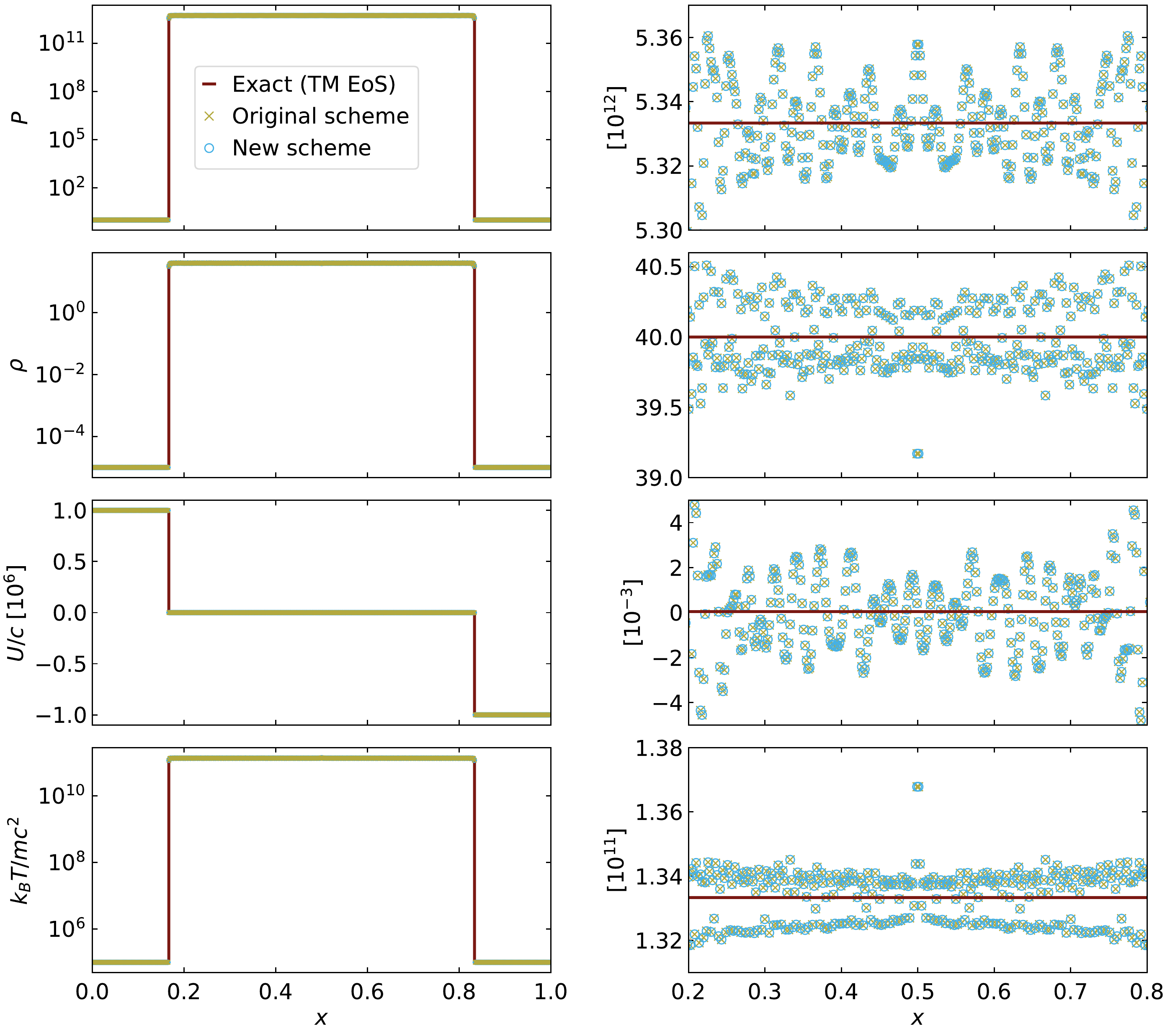}
\caption{Riemann problem in the UR limit with a head-on collision of two identical gases with $\beta\gamma=10^{6}$ and $k_{B}T/mc^2=10^{5}$ at $t=1.0$. The left column shows the entire simulation domain, while the right column shows the zoom-in image of the post-shock region. From top to bottom: pressure, proper mass density, four-velocity, and temperature. Note that we plot the Mach number in the zoom-in image (right column) of four-velocity so as to readily compare the amplitude of velocity oscillation with sound speed. The new scheme ($\tikz\draw[HeadOnCollisionNew, fill=white,line width=1] (0,0) circle (.5ex);$) fully agrees with the original scheme ($\MyCross[draw=HeadOnCollisionOri,fill=HeadOnCollisionOri]$) not only on the large-scale profiles but also on the small-scale errors, meaning that the new scheme ($\tikz\draw[HeadOnCollisionNew,fill=white,line width=1] (0,0) circle (.5ex);$) does not sacrifice (or improve) the numerical accuracy in the UR limit for  that in the NR limit.}
\label{fig:head-on collision shock tube}
\end{figure*}

We simulate a head-on collision of two identical gases with $\gamma=10^{6}$ and $k_{B}T/mc^2=10^5$ with uniform 512 grids. The computational domain is in the interval $[0,1]$. The initial discontinuity is located at $x=0.5$. The first row of Table~\ref{tb:IC_RiemannProblems} presents the initial right and left states. \Cref{fig:head-on collision shock tube} shows the results at $t=1.0$. The left panels show the entire simulation domain, while the right panels show the zoom-in image of the post-shock region, which has been violently heated up to ultra-relativistically hot temperature ($k_{B}T/mc^2 \sim 10^{11}$) by the extremely high-$\gamma$ gases flowing inwards from both sides. As can be seen, the new scheme ($\tikz\draw[HeadOnCollisionNew,fill=white,line width=1] (0,0) circle (.5ex);$) fully agrees with the original scheme ($\MyCross[draw=HeadOnCollisionOri,fill=HeadOnCollisionOri]$) on the large-scale profile but also on the small-scale errors, meaning that the new scheme does not sacrifice the numerical accuracy in the UR limit. In addition, we notice that the non-negligible and spurious waves occur in the post-shock region, which are not due to root-finding iterations but to spatial discretization errors as the spurious waves can be reduced by increasing spatial resolution.

\definecolor{MixedLimitsOri}{RGB}{0,0,0}
\definecolor{MixedLimitsExact}{RGB}{224, 168, 0}
\definecolor{MixedLimitsNew}{RGB}{255, 20, 0}
\definecolor{MixedLimitsExactNR}{RGB}{0, 21, 255}
\definecolor{MixedLimitsExactUR}{RGB}{0, 166, 28}

\subsubsection{Mixed limits}
\label{subsection:NRLimit}
To demonstrate that the new scheme can handle a large dynamical range covering both extremely hot and extremely cold gases, we simulate a nontrivial Riemann problem where the temperature straddles between the high- and low-$T$ limits. This initial condition evolves into a cold left-traveling rarefaction wave separated by a contact discontinuity to match an extremely hot downstream of an ultra-relativistic shock traveling toward the right. Also, we have numerically derived the exact solution of this particular Riemann problem with the TM EoS (see Appendix \ref{appendix:exact solution}). The second row in \Cref{tb:IC_RiemannProblems} shows the initial left and right states. The simulation adopts a computational domain [0,100] with 102,400 cells. Since the speed of the right-traveling shock is 276 times faster than that of the left-traveling rarefaction wave, we put the initial discontinuity at $x=5\times10^{-2}$ to provide an ample space for the right-traveling shock. 

\begin{figure*}
\includegraphics[width=\linewidth]{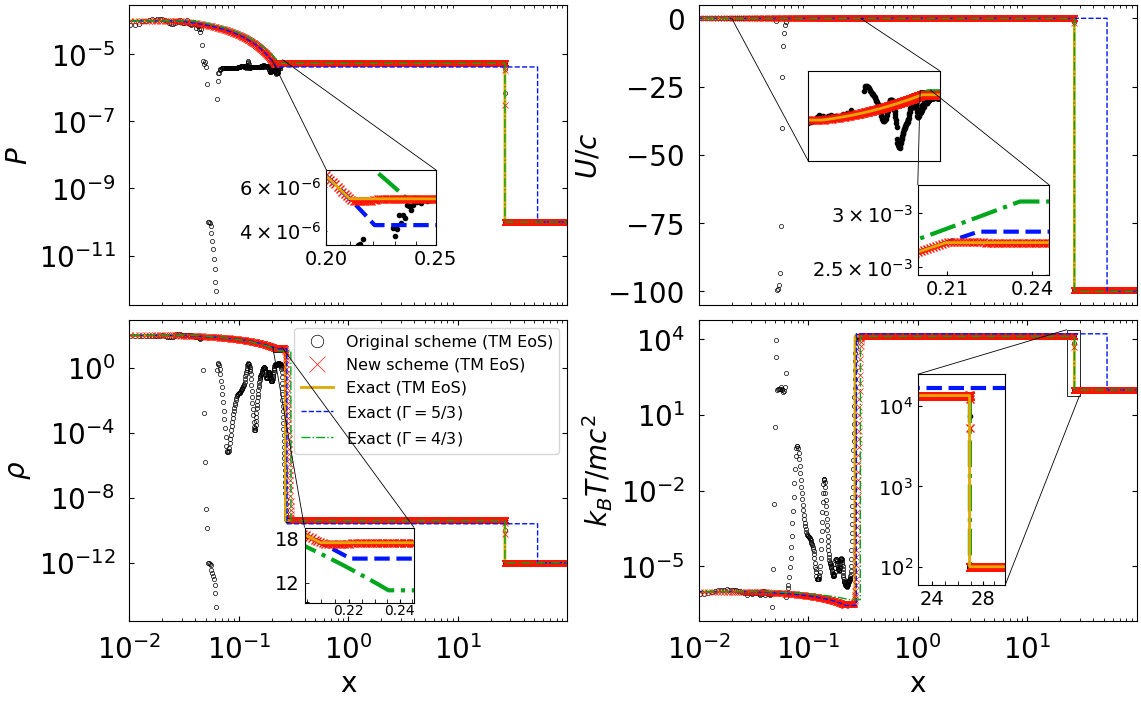}
\centering
\caption{Riemann problem in the mixed UR and NR limits at $t=80$. The second row in Table \ref{tb:IC_RiemannProblems} shows the initial condition. Clock-wise from top-left: pressure, 4-velocity, temperature, and proper mass density. We find not only that the shock front at $x=26$ is well resolved by 3--4 cells but also that the new scheme ($\MyCross[draw=MixedLimitsNew,fill=MixedLimitsNew]$) agrees well with the exact solution of the TM EoS ($\MySolidLine[draw=MixedLimitsExact,fill=MixedLimitsExact]$), as shown in all insets. The L1 error of the density profile from the original scheme ($\tikz\draw[MixedLimitsOri,fill=white,line width=1] (0,0) circle (.5ex);$) is 23 per cent within the region between the head of rarefaction wave and initial discontinuity (i.e. $2.67\times 10^{-2}<x<5\times10^{-2}$), consistent with the 20 per cent error estimated by \Cref{eq:error in the NR limit} with $k_{B}T/mc^2=8\times10^{-6}$. However, in the region $5\times10^{-2}<x<0.3$ swept by the right-traveling contact discontinuity, errors of the original scheme are much larger than the estimate, which requires further investigation. The TM profiles match well with both the $\Gamma=5/3$ profiles ($\MyDashedLine[draw=MixedLimitsExactNR,fill=MixedLimitsExactNR]$) in the NR region ($x<0.21$) and the $\Gamma=4/3$ profiles ($\MyDashedDottedLine[draw=MixedLimitsExactUR,fill=MixedLimitsExactUR]$) in the UR region ($x>0.27$)}.
\label{fig:non-relativistic shock tube}
\end{figure*}

\Cref{fig:non-relativistic shock tube} shows the results at $t=80$, where there are three points to be emphasized. First, we find not only that the shock front at $x=26$ is well resolved by 3--4 cells but also that the new scheme ($\MyCross[draw=MixedLimitsNew,fill=MixedLimitsNew]$) agrees well with the exact solution of the TM EoS ($\MySolidLine[draw=MixedLimitsExact,fill=MixedLimitsExact]$), as shown in all insets. Second, the L1 error, defined by \Cref{relative L1 error}, of the density profile from the original scheme ($\tikz\draw[MixedLimitsOri,fill=white,line width=1] (0,0) circle (.5ex);$) is 23 per cent within the region between the head of rarefaction wave ($x=2.67\times 10^{-2}$, the third number from top in the leftmost column of Table \ref{tb:exact solution} in Appendix \ref{appendix:exact solution}) and initial discontinuity ($x=5\times10^{-2}$), consistent with the 20 per cent error estimated by \Cref{eq:error in the NR limit} with $k_{B}T/mc^2=8\times10^{-6}$. Similar conclusions can be drawn for other physical quantities. However, in the region $5\times10^{-2}<x<0.3$ swept by the right-traveling contact discontinuity, errors of the original scheme are much larger than the estimate, which requires further investigation. Third, the solutions of the TM EoS ($\MySolidLine[draw=MixedLimitsExact,fill=MixedLimitsExact]$) match well with both $\Gamma=5/3$ ($\MyDashedLine[draw=MixedLimitsExactNR,fill=MixedLimitsExactNR]$) in the NR region ($x<0.21$) and $\Gamma=4/3$ ($\MyDashedDottedLine[draw=MixedLimitsExactUR,fill=MixedLimitsExactUR]$) in the UR region ($x>0.27$). It demonstrates the capability of capturing the transition from $\Gamma=5/3$ (for $k_{B}T/mc^2 \rightarrow 0$) to $\Gamma=4/3$ (for $k_{B}T/mc^2 \rightarrow \infty$) for the new scheme. The exact solutions of this test are shown in \Cref{tb:exact solution} in Appendix \ref{appendix:exact solution}.
 
\subsection{Multi-dimensional grid effects for high-$\mathscr{M}$ flows}
\label{Multi-dimensional grid effects}
To investigate the detrimental impact of grid effects on the evolution of ultra-relativistic and high Mach number hydrodynamic problems, we separately simulate two identical three-dimensional mono-direction flow with different flow directions. One flow is along the diagonal direction of the simulation box and the other is parallel to the grid direction. Both simulations share the same numerical set-up as follows. Flows are initially represented by cylinders extending to the boundaries of a periodic cubic box with a width $L$. The cylinder diameter is $D=0.028L$. The proper mass density ratio of the flow and the ambient is $\rho_{\text{flow}}/\rho_{\text{amb}}=10^{-5}$. The temperatures of the flows and the ambient are $k_{B}T_{\text{flow}}/mc^2=1.0$ and $10^{-5}$, respectively. The four-velocity ($\gamma \beta$) profile inside the flow source is $10^{6}\left(1+\cos{\left(2\pi r/D\right)}\right)$, where $r$ is the distance from the flow axis inside the source. Other physical quantities are uniformly distributed inside the source.

The AMR base level is covered by $64^{3}$ cells in all cases. We adopt the gradient of the proper mass density and the magnitude of $\abs{\mathbf{M}}/D$ as the two inclusive refinement criteria. We refine a patch if the gradient of a cell satisfies
\begin{equation}
\frac{\Delta h_{\ell}}{Q}
\left(
\abs{\frac{\partial Q}{\partial x}}+
\abs{\frac{\partial Q}{\partial y}}+
\abs{\frac{\partial Q}{\partial z}}
\right)
> C_{Q},
\label{threshold for gradient}
\end{equation}
where $Q=\rho$, $C_{Q}=0.3$, and $\Delta h_{\ell}$ is the cell size at refinement level $\ell$. This criterion aims to capture the finger structure due to instabilities at the interface between the flow and the ambient gases. Also, a patch will be refined when any cell satisfies $\abs{\mathbf{M}}/D>10^{4}$ so that the high-speed region is refined to the finest level.

\begin{figure*}
  \centering
  \subfigure[]
    {
      \includegraphics[width=0.45\linewidth]{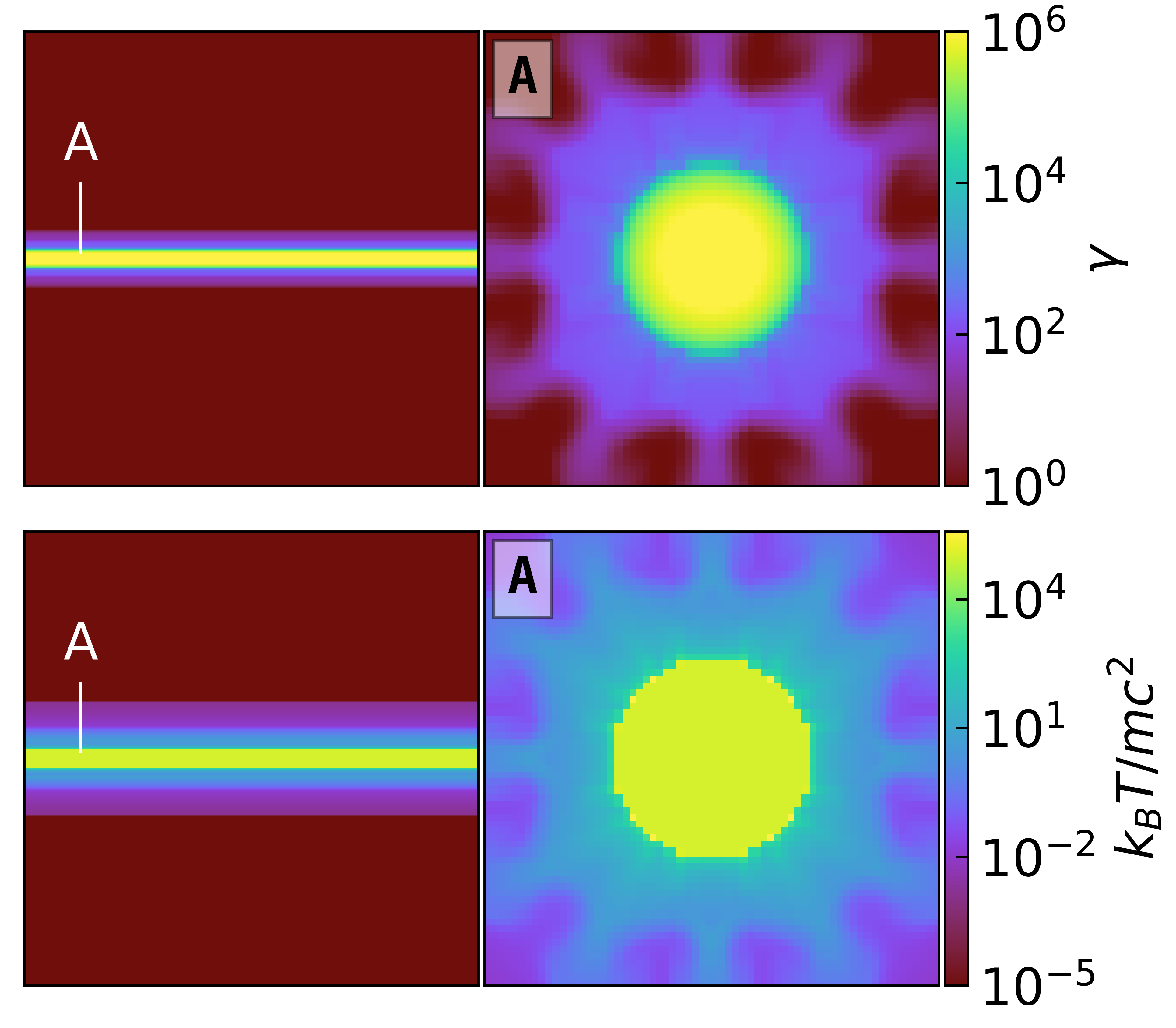}
      \label{fig:AAXX} 
    }
  \subfigure[]
    {
      \includegraphics[width=0.45\linewidth]{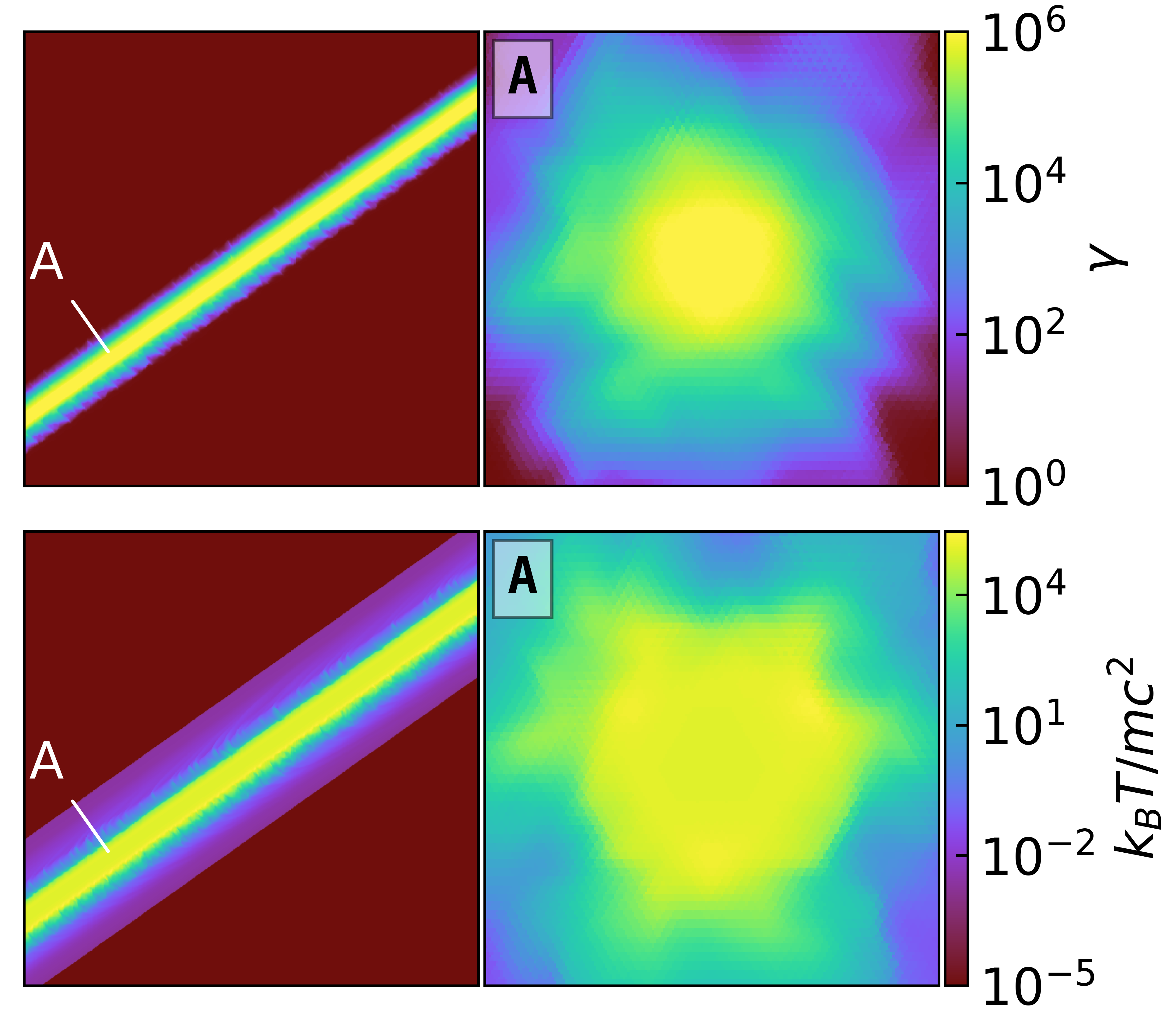}
      \label{fig:DiagonalFlowLow} 
    }
  \subfigure[]
    {
      \includegraphics[width=0.45\linewidth]{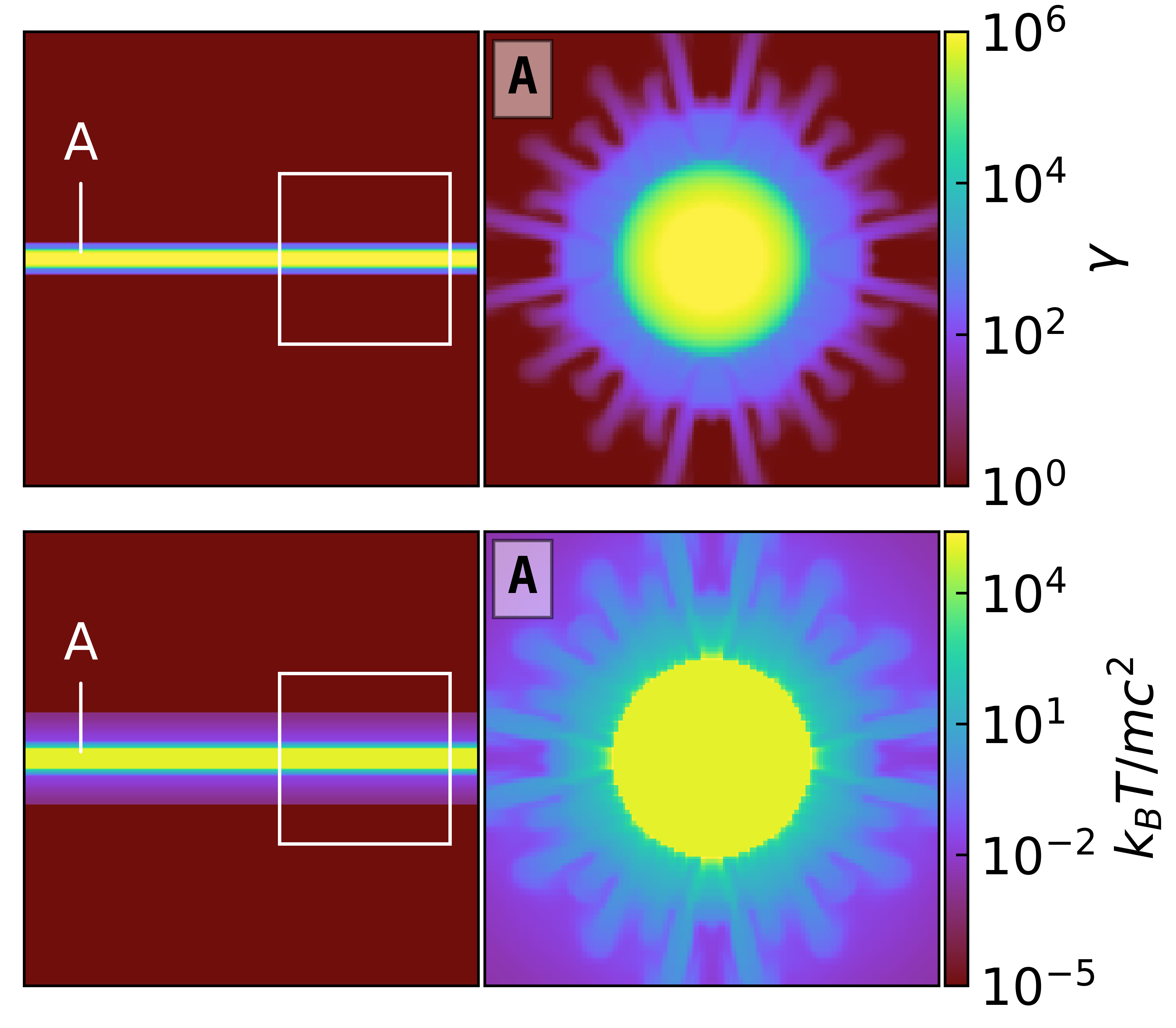}
      \label{fig:HorizontalFlowHig} 
    }
  \subfigure[]
    {
      \includegraphics[width=0.45\linewidth]{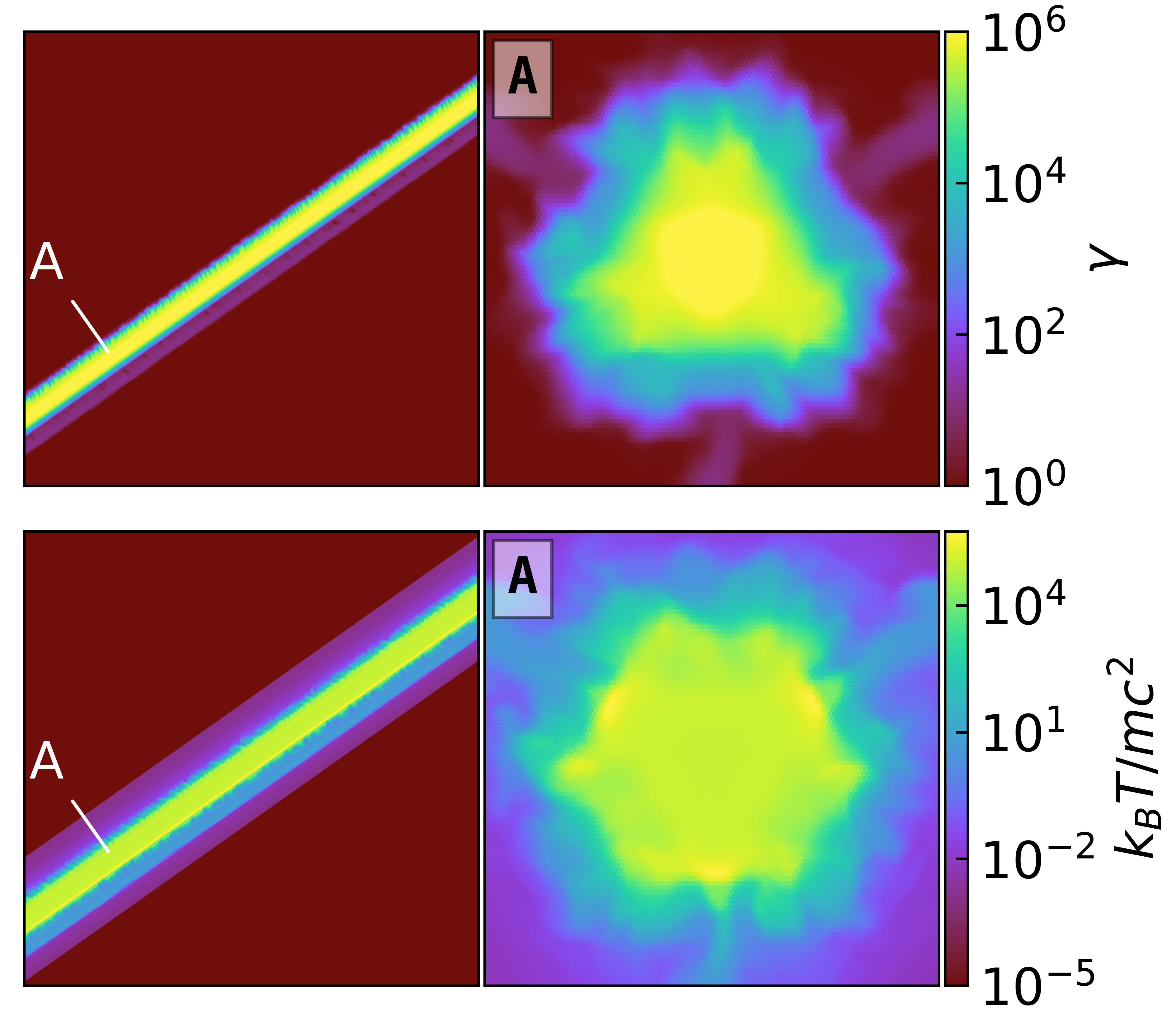}
      \label{fig:DiagonalFlowHig} 
    }
  \caption{
Ultra-relativistic flows propagating along different spatial directions with respect to the grids. In all subfigures (a)--(d), the left and right columns are the longitudinal and transverse slices, respectively. Longitudinal slices are taken through the flow source while the transverse slices are taken through the label `A'. The flow diameter is resolved by 28 cells in (a) and (b) and by 56 cells in (c) and (d). The flow has an extremely high Mach number ($\mathscr{M}\sim 10^{6}$) leaving any instability short of time to develop, and one expects a smooth flow-ambient interface. However, the fuzzy-looking cross-sections in the transverse slices of the oblique flow (right columns in (b) and (d)) suggest that the high-speed flow induces false instability when the flow travels obliquely across Cartesian grids. Increasing the spatial and temporal resolution does not help the numerical solution of an oblique flow to converge to that of a parallel flow.}
  \label{fig:GridEffect}
\end{figure*}

\Cref{fig:GridEffect} shows the simulation results at $t=0.4L/c$. In Figures \ref{fig:AAXX} and \ref{fig:DiagonalFlowLow}, we adopt four AMR levels to ensure that the flow diameter can be resolved by 28 cells. The extremely high Mach number ($\mathscr{M}\sim 10^{6}$) flow leaves any instability short of time to develop, and one expects a smooth flow-ambient interface. However, the interface of the oblique flow turns out to be subject to severe dissipation. The fuzzy-looking cross-sections in the transverse slices of the oblique flow (right column in \Cref{fig:DiagonalFlowLow}) suggest that the dissipation is caused by numerical instabilities when high Mach number flow travels obliquely across Cartesian grids. This numerical problem is not limited to relativistic high Mach number flows but also occurs in non-relativistic high Mach number flows.

To examine this issue further, we increase the spatial resolution by a factor of 2 and decrease the time-step by a factor of 0.3 from the standard Courant condition. The results (\Cref{fig:HorizontalFlowHig,fig:DiagonalFlowHig}) indicate that increasing spatial and temporal resolution can neither significantly ameliorate the dissipation nor help the oblique flow converge to the horizontal flow. This artificial grid effect can adversely influence the study of high-speed jets, especially for hydrodynamical instabilities near the jet boundaries.

\begin{figure}
\includegraphics[width=\columnwidth]{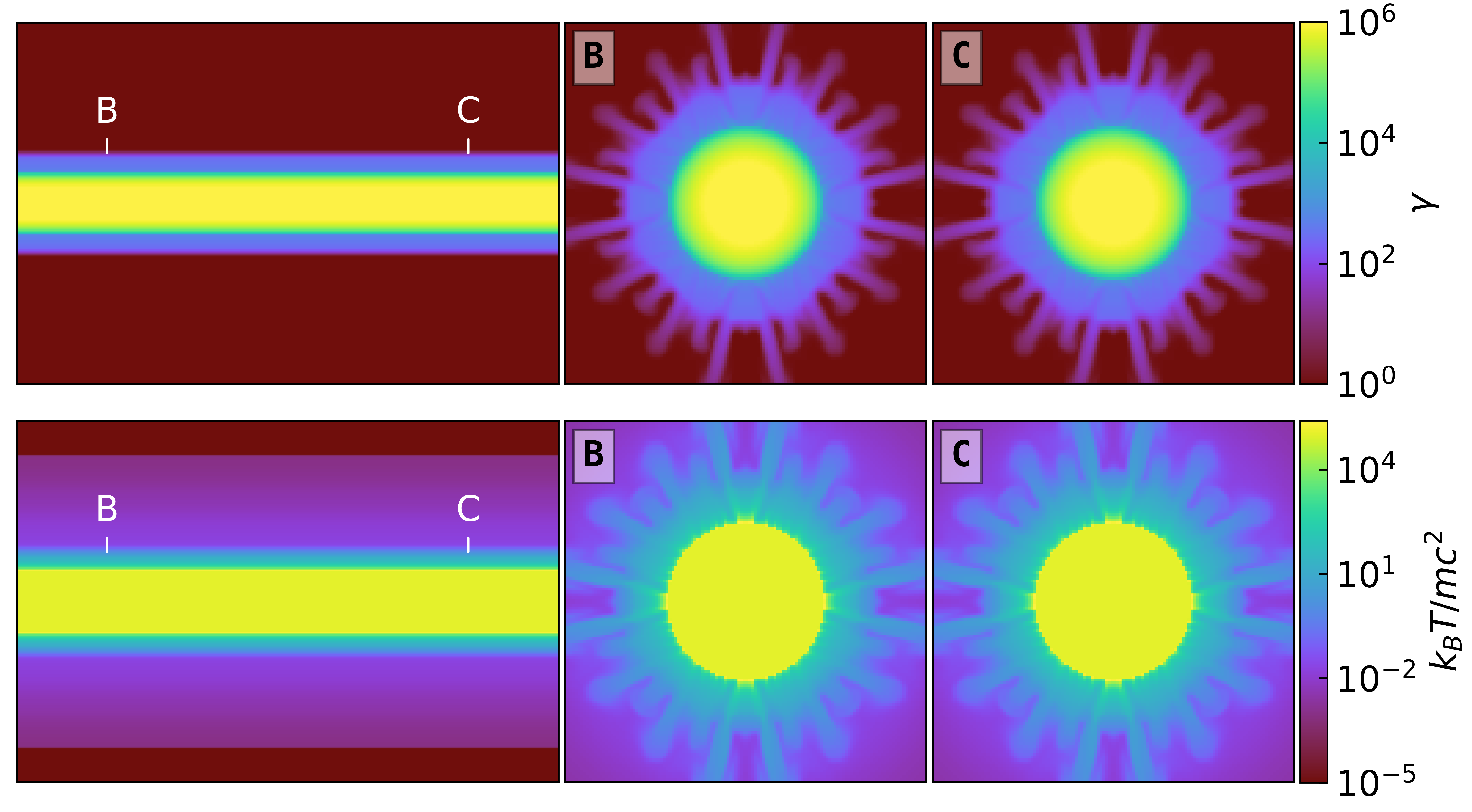}
\centering
\caption{Close-up view of the rectangular region in \Cref{fig:HorizontalFlowHig} with longitudinal (first column from the left) and transverse (other columns) slices passing through `B' and `C', respectively. Compared with \Cref{fig:HorizontalFlowHig}, it clearly shows that the finger pattern consists of 2-D flat sheets along the flow.}
\label{fig:HorizontalFlowHigZoomIn}
\end{figure}

An example of this boundary instability is the finger-like pattern observed immediately outside the parallel flow (right column in \Cref{fig:AAXX}), which we believe to arise from a genuine instability seeded by discritization noise. The finger-like pattern has a higher temperature than the ambient, and in fact consists of two-dimensional flat sheets along the flow. This is demonstrated in \Cref{fig:HorizontalFlowHigZoomIn} with transverse slices cut through `B' and `C'. The patterns are identical to that cut through `A' in \Cref{fig:HorizontalFlowHig}. These 2-D sheet pattern persists even after adding 1 per cent level of white noise into the background density, illustrating that the coherence of sheets along the flow direction is genuinely generated by the high-speed flow boundaries. This finger pattern is similar, but not identical, to the curvature-driven fingers of a knotted jets reported recently \citep{Gourgouliatos2017}. Our flow has a smooth and parallel boundary without any curvature to drive the fingers.

\section{Performance scaling}
\label{Performance scaling}
We measure both strong and weak scalings of $\textsc{gamer-sr}$ with AMR and hybrid MPI/OpenMP/GPU
parallelization. The simulations were conducted on the \texttt{Piz-Daint} supercomputer that provides a 12-core Intel Xeon E5-2690 CPU and a Tesla P100 GPU on each computing node. Strong and weak scalings are defined as how the simulation wall time varies with the number of computing nodes for a fixed total problem size and for a fixed problem size per node, respectively. We launch one MPI process with 12 OpenMP threads per node and enable GPU acceleration with single precision.

\begin{figure}
\includegraphics[width=\columnwidth]{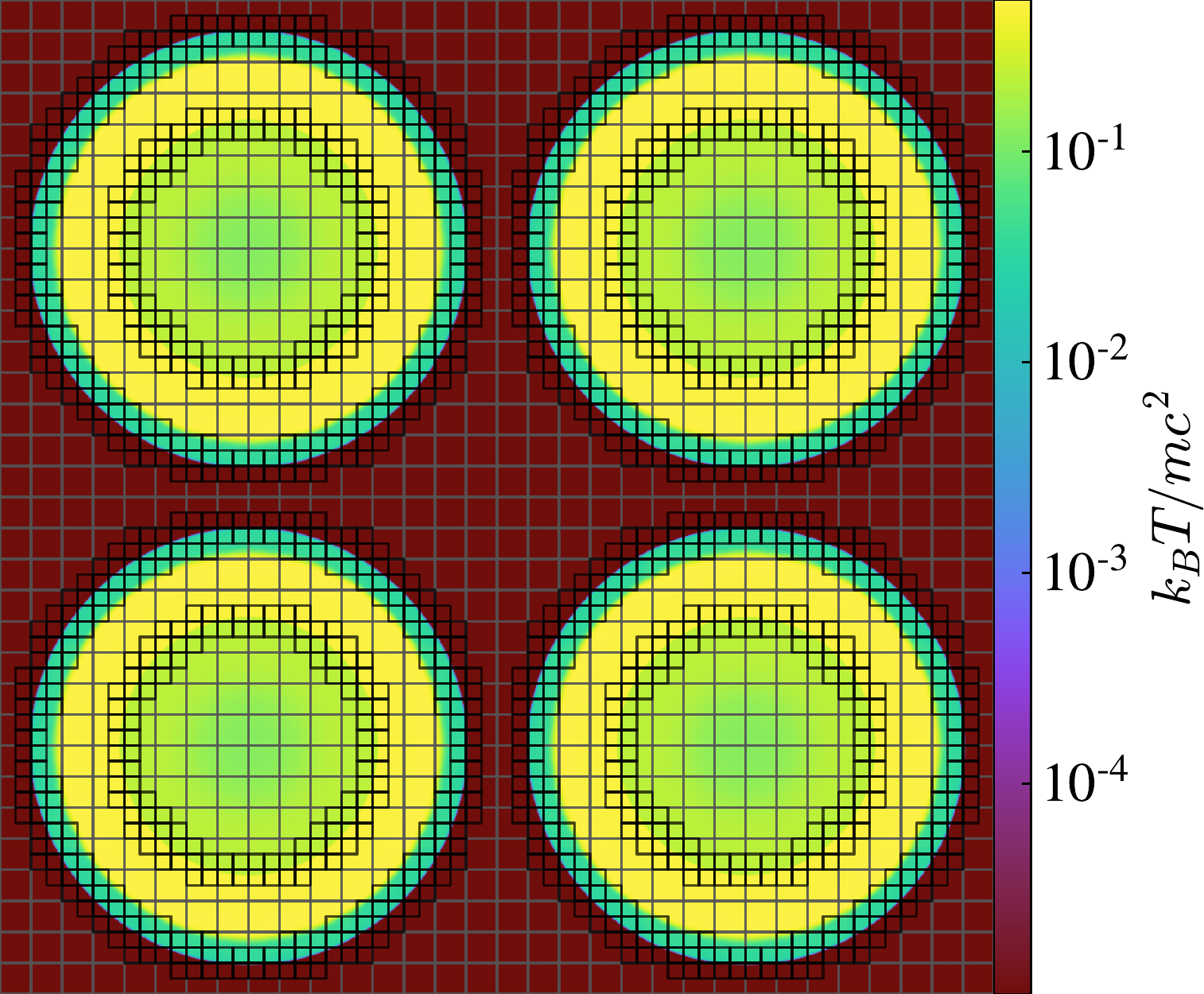}
\centering
\caption{Temperature slice through the centre of blast waves at $t=0.5L/c$, with the grid patches overlaid in the case of 8 nodes in the weak scaling test.}
\label{fig:multi-blast waves}
\end{figure}

We divide this section into two parts. First, we measure the strong scaling of a relativistic jet simulation. The simulation set-up, such as initial condition, boundary condition, and grid refinement, follows those described in Section \ref{Limb-brightened jet}. Second, we present the weak scaling for periodic and spherical multi-blast waves test (see \Cref{fig:multi-blast waves}).

\begin{figure}
\includegraphics[width=\columnwidth]{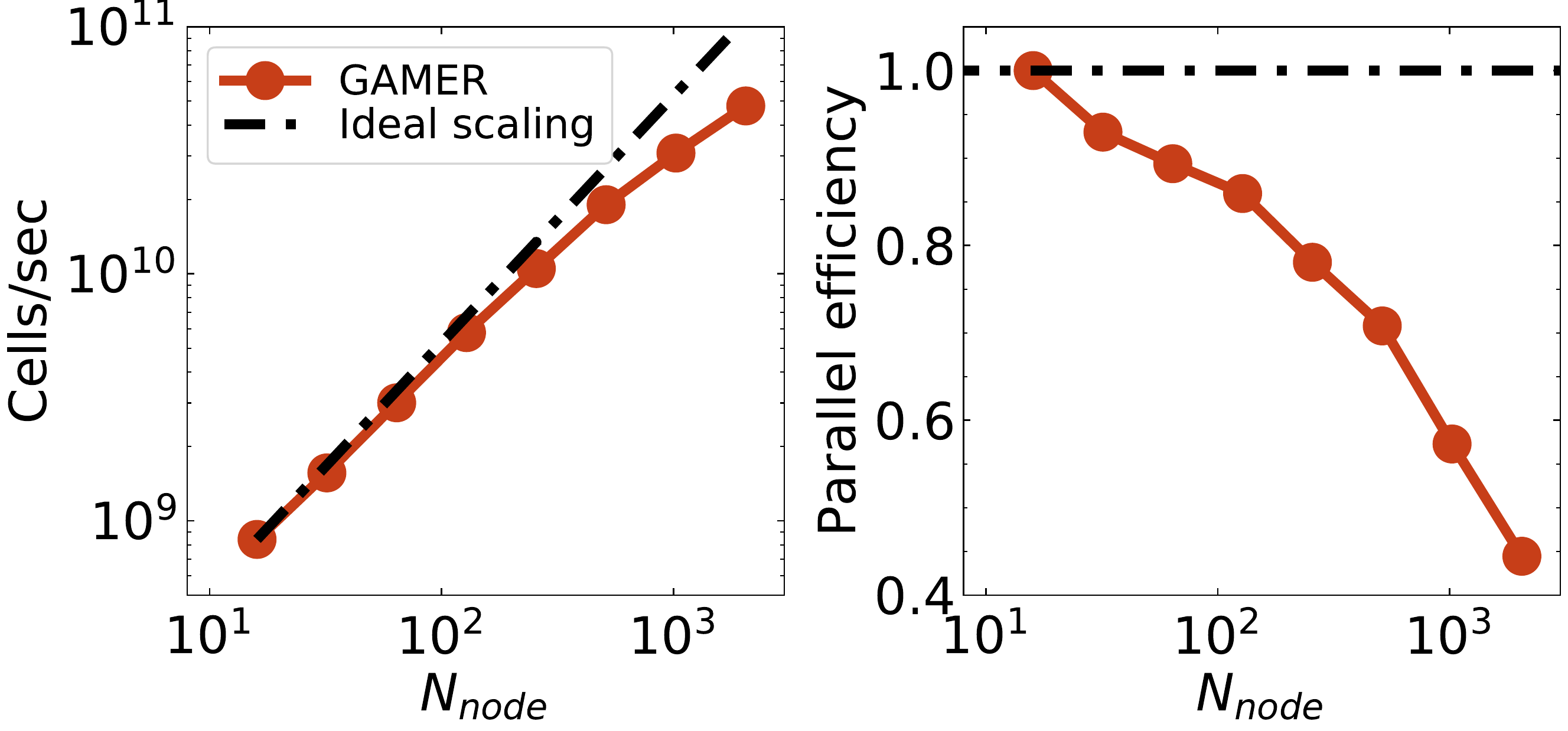}
\caption{Strong scaling using 1--2048 GPU nodes (left panel: cell updates per second; right panel: parallel efficiency). The deviation from the ideal scaling is mainly due to MPI communication, the time fraction of which increases by a factor of 10 when increasing the number of nodes from 64 to 2048.}
\label{fig:strong scaling}
\end{figure}

    \emph{(1) Strong scaling:}\\
    \Cref{fig:strong scaling} shows the strong scaling results. The parallel efficiency for strong scaling is defined by $\left[T\left(N_{\text{ref}}\right)/T\left(N_{\text{node}}\right)\right]/\left(N_{\text{node}}/N_{\text{ref}}\right)$, where  $T(N_{\text{node}})$ is the simulation wall time using $N_{\text{node}}$ nodes. $N_{\text{ref}}$ is the number of nodes for reference and is fixed to 16 in our test. The overall performance reaches $5\times 10^{10}$ cell updates per second with 2048 GPU nodes, corresponding to a parallel efficiency of $\sim$ 45 per cent. The deviation from the ideal scaling is mainly due to MPI communication, the time fraction of which increases by a factor of 10 when $N_{\text{node}}$ increases from 64 to 2048.

    \emph{(2) Weak scaling:}\\
    The periodic computational domain is composed of identical cubic subdomain, each of which has a volume of $L^3$ and has an explosion source at its own centre with a radius of $r_{\text{src}}=0.4L$ and an ultra-relativistic temperature of $k_{B}T_{\text{src}}/mc^2=10^{5}$. The uniform ambient gas has a non-relativistic temperature of $k_{B}T_{\text{amb}}/mc^2=10^{-5}$ and a density of $\rho_{\text{src}}=\rho_{\text{amb}}=1.0$. Each subdomain is composed of a $64^3$ base-level grid with three refinement levels, where we refine patches based on the gradient of the reduced energy density. All blast waves evolve from $t=0$ to $t=0.5L/c$. We measure the overall performance and parallel efficiency using $1-2048$ nodes, where each node computes one subdomain. \Cref{fig:multi-blast waves} shows a temperature slice ($z=1.5L$) through the centre of four blast waves at $t=0.5L/c$, with the grid patches overlaid.

\begin{figure}
\includegraphics[width=\columnwidth]{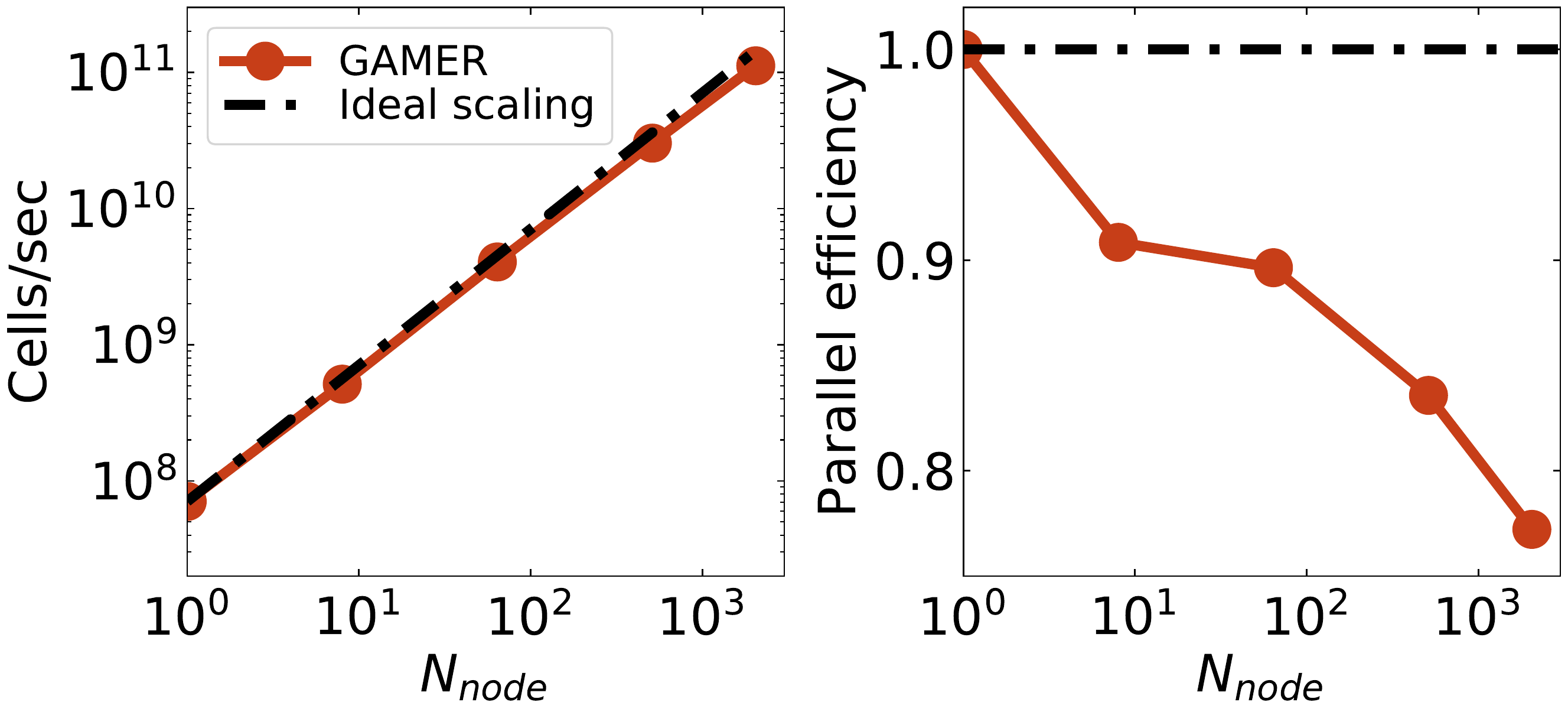}
\caption{Weak scaling using 1--2048 GPU nodes (left panel: cell updates per second; right panel: parallel efficiency). The parallel efficiency drops from 0.90 to 0.78 when $N_{\text{node}}$ increases from 64 to 2048 mainly because the MPI time fraction increases from 10.3 to 18.8 per cent.}
\label{fig:weak scaling}
\end{figure}

\Cref{fig:weak scaling} shows the weak scaling results. The parallel efficiency for weak scaling is defined by $T(1)/T(N_{\text{node}})$, where $T(N_{\text{node}})$ is defined the same as the strong scaling. The parallel efficiency is measured to be 90 per cent with 64 nodes and 78 per cent with 2048 nodes, achieving a peak overall performance of $1.3\times 10^{11}$ cell updates per second with 2048 nodes. The drop of parallel efficiency is mainly due to MPI communication, the time fraction of which increases from 10.3 ($N_{\text{node}}=64$) to 18.8 per cent ($N_{\text{node}}=2048$).

We remark that the strong and weak scaling tests demonstrate 55 and 80 per cent parallel efficiencies, respectively, with 1024 nodes on the \texttt{Piz-Daint} supercomputer. Moreover, the peak performance on a single Tesla P100 GPU achieves $7\times 10^{7}$ cell updates per second, which is about one-third of the peak performance of non-relativistic hydrodynamics \citep{gamer-2}.

\definecolor{BlastSphOld}{RGB}{88, 64, 230}
\definecolor{BlastTriOld}{RGB}{219, 186, 0}
\definecolor{BlastSphNew}{RGB}{199, 62, 24}
\definecolor{BlastTriNew}{RGB}{59, 173, 130}

\section{Astrophysical Applications}
\label{Astrophysical Applications}
\subsection{Triaxial blast wave}\label{Ellipsoid blast wave test}
This triaxial relativistic blast wave problem models a hypothetical astrophysical mega-explosion driven by an ultra-relativistically hot plasma source absent of particular symmetry. It is an atypical test for which we verify the code's ability to deal with strong 3D shocks. The simulation evolves a blast wave from a triaxial source in a homogeneous medium. The triaxial source has aspect ratios $1:1.5:2$ with a semi-major axis $0.01L$ aligned with the diagonal direction, where $L$ is the width of a cubic computational box. The source is filled with a uniform ultra-relativistic ($k_{B}T_{\text{src}}/mc^2=10^{6}$) plasma and the ambient is filled with a uniform non-relativistic ($k_{B}T_{\text{amb}}/mc^2=10^{-9}$) HII gas. The density is homogeneous throughout the entire domain with $\rho_{\text{src}}=\rho_{\text{amb}}=1.0$. After the system quickly relaxes, the hot plasma rapidly expands driving a forward shock traveling almost at the speed of light.

The AMR base level is covered by $32^3$ cells with the periodic boundary condition. The highest refinement level is 9 so that the initial source can be adequately resolved by approximately 82 cells along the minor axis. To refine both the initial source and the thin shell of the blast wave shock, we adopt the gradient of the reduced energy density as the refinement criterion, with $Q=\tilde{E}$ and $C_{Q}=1.0$ in \Cref{threshold for gradient}.

For comparison, we also simulate a spherical blast wave to understand how the initially triaxial shape affects the evolution of the ultra-relativistic blast wave. Both the spherical and triaxial cases have the same simulation set-up and the same source volume, i.e. $r/L=\sqrt[3]{0.01\times0.0075\times0.005}$, where $r$ is the radius of the spherical source.

\begin{figure*}
\includegraphics[width=\linewidth]{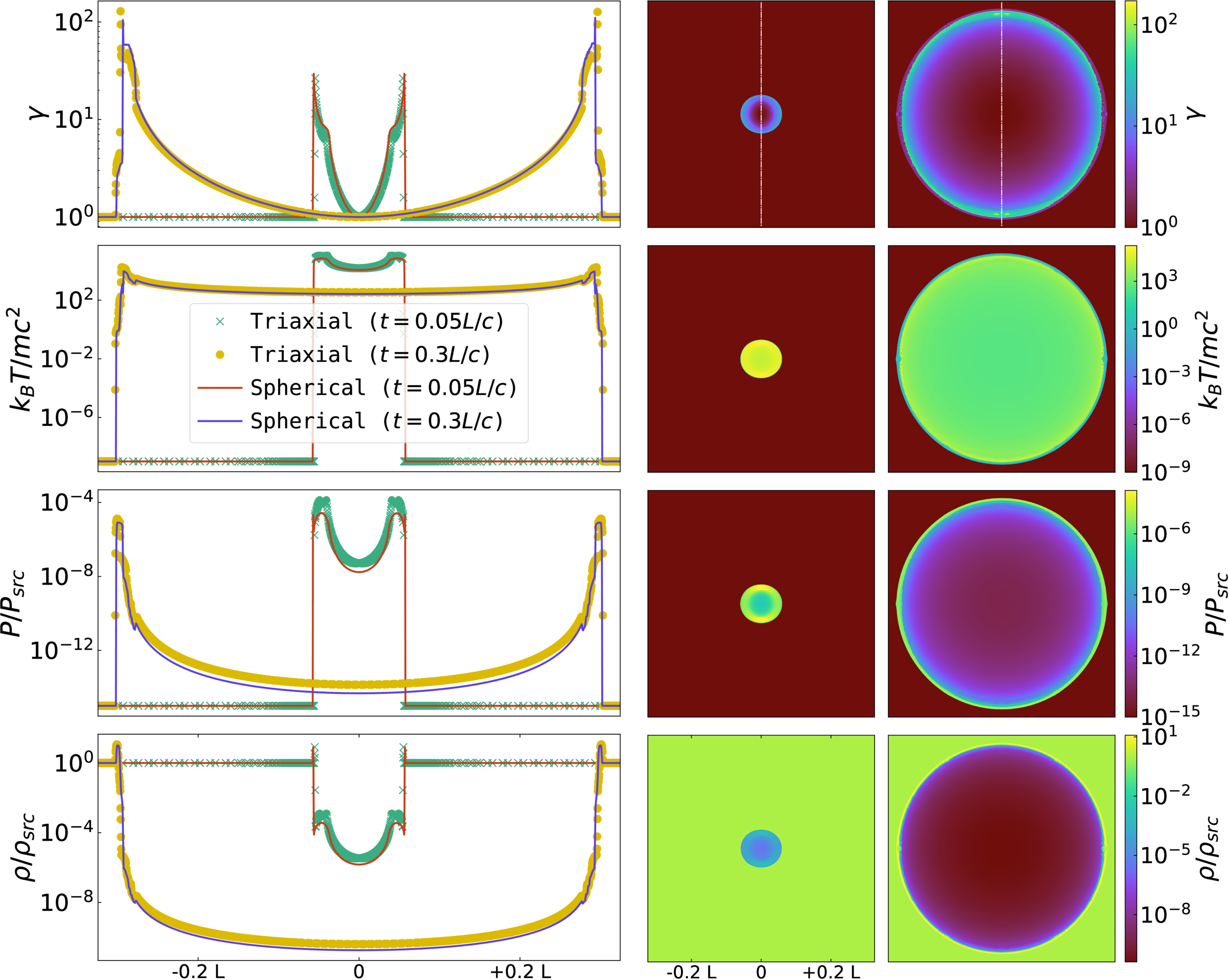}
\centering
\caption{Triaxial blast wave test. The middle ($t=0.05L/c$) and right ($t=0.3L/c$) columns show the slices passing through the triaxial source at the mid-plane of its intermediate axis (i.e. the horizontal and vertical axes are along the major and minor axes, respectively). The left column shows the profiles along the minor-axis (i.e. the white dotted-dashed line in the $\gamma$ map).}
\label{fig:Ellipsoid blast wave}
\end{figure*}

\Cref{fig:Ellipsoid blast wave} shows the results. We observe that the interior hot plasma pushes out a contact discontinuity immediately inward of the shock and that the thickness of the shell between the contact discontinuity and the shock diminishes in time. In early time, the triaxial profiles ($\MyCross[draw=BlastTriNew,fill=BlastTriNew]$) at $t=0.05L/c$ deviate from the spherical counterparts ($\MySolidLine[draw=BlastSphNew,fill=BlastSphNew]$), especially in the pressure and proper mass density, although the shock positions almost coincide. However, at a later time, the profiles at $t=0.3L/c$ show no significant difference between the triaxial ($\tikz\draw[BlastTriOld,fill=BlastTriOld,line width=1] (0,0) circle (.5ex);$) and spherical ($\MySolidLine[draw=BlastSphOld,fill=BlastSphOld]$) blast waves, indicating the initial shape of the source does not have a great impact on the asymptotic evolution of ultra-relativistic blast waves. 

\begin{figure}
\includegraphics[width=\linewidth]{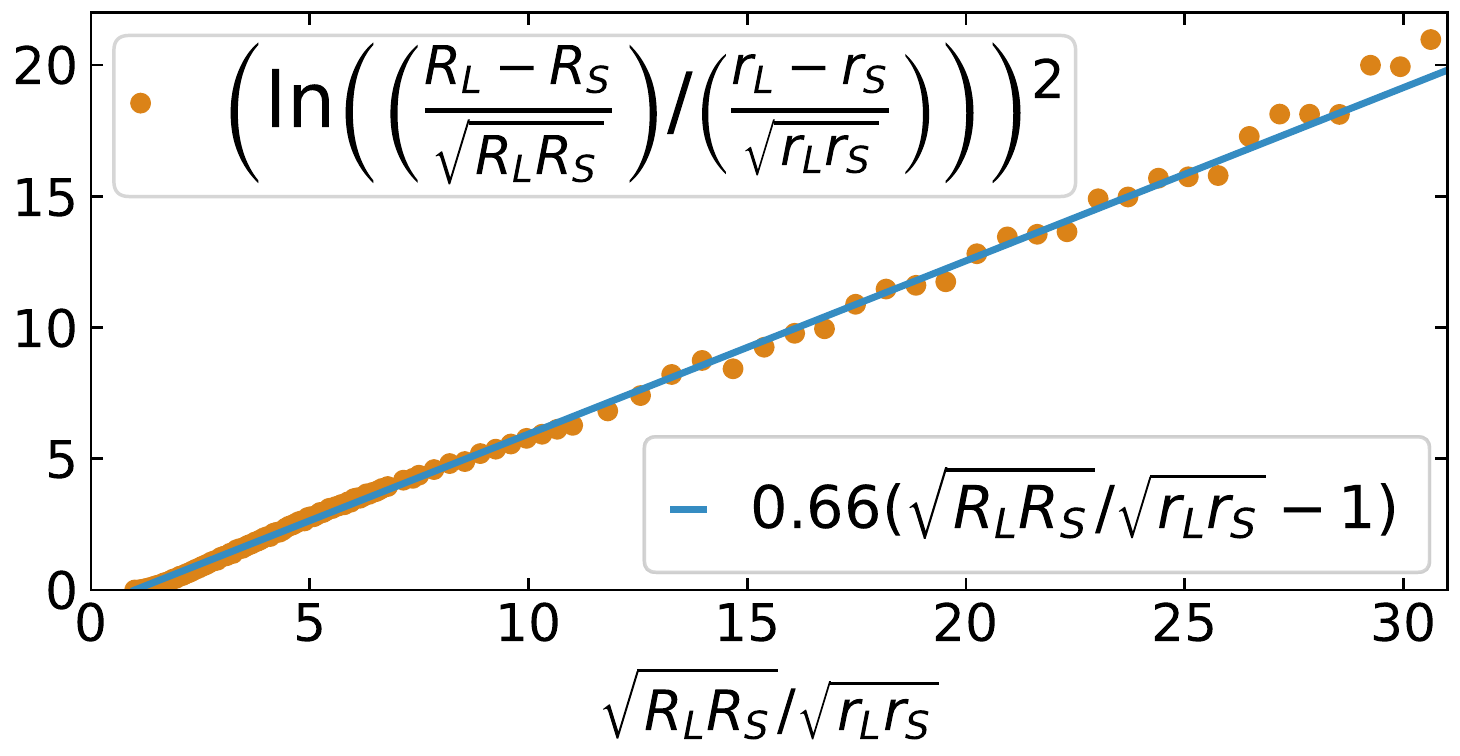}
\caption{Damping of the triaxiality in the triaxial blast wave test, where $R_{L}$ and $R_{S}$ are the radii of the triaxial blast wave along the semi-major ($r_{L}$) and semi-minor axes ($r_{S}$) of the initial source.}
\label{fig:BlastFitting}
\end{figure}

To further investigate how the triaxial blast wave evolves into a spherical one, we extract the radii $R_{L}(t)$ and $R_{S}(t)$ of the triaxial blast wave along the semi-major ($r_{L}$) and semi-minor axes ($r_{S}$) of the initial source from simulation data. As shown in \Cref{fig:BlastFitting}, we find that the dimensionless quantity $\left(\ln \left(\left(\frac{R_{L}-R_{S}}{\sqrt{R_{L}R_{S}}}\right)/ \left(\frac{r_{L}-r_{S}}{\sqrt{r_{L}r_{S}}}\right)\right)\right)^{2}$ is approximately equal to $0.66\left(\sqrt{R_{L}R_{S}}/\sqrt{r_{L}r_{S}}-1\right)$. This dependence suggests that the triaxiality is damped out with the blast wave propagation by the relation:
\begin{equation}
\begin{aligned}
          &\frac{R_{L}(t)-R_{S}(t)}{\sqrt{R_{L}(t)R_{S}(t)}} =\\ &\left(\frac{r_{L}-r_{S}}{\sqrt{r_{L}r_{S}}}\right)\text{exp}\left[-0.81\left(\frac{\sqrt{R_{L}(t)R_{S}(t)}}{\sqrt{r_{L}r_{S}}}-1\right)^{1/2}\right].
\end{aligned}
\label{eq:FittingBlast}
\end{equation}
 
We remark that this test problem also demonstrates that \textsc{gamer-sr} can successfully handle ultra-relativistic gases embedded in a cold HII region, which can be difficult for conventional SRHD codes.

\subsection{Limb-brightened jet}
\label{Limb-brightened jet}
Most active galactic nuclei (AGN) jets in VLBI observations appear ridge-brightened, while limb-brightened jets are rare and have been reported only in a few nearby radio galaxies, such as Mrk 501 \citep{Giroletti2004}, M87 (\citealt{Asada2012}; \citealt{Kim2018}), Cygnus A \citep{Boccardi2015}, and 3C84 (\citealt{Nagai_2014}; \citealt{Giovannini2018}). Motivated by these observations, we simulate a three-dimensional SRHD jet using \textsc{gamer-sr} to study its acceleration and collimation in the hope to shed light on the limb-brightened jets.

We adopt the gradient of the reduced energy density and the magnitude of $\abs{\mathbf{M}}/D$ as the two inclusive refinement criteria. A patch is refined if any cell satisfies either \Cref{threshold for gradient} with $Q=\tilde{E}$ and $C_{Q}=0.1$ or $\abs{\mathbf{M}}/D>20$. The first criterion aims to capture the strong terminal shock and the cocoon, while the second one ensures that the central `spine' region of the jet can be properly resolved.

The jet is continuously ejected from a cylindrical source with four-velocity $\beta\gamma=10.0$ ($\gamma \sim 10.05$). The proper mass density ratio between the jet source and the ambient gases is set to 1.0. The temperature ($k_{B}T/mc^2$) of the source and the ambient gases are set to 0.5 and $10^{-5}$, respectively. The outflow is thus an extremely under-pressured jet. Both the diameter and the length of the cylindrical source are well resolved by 32 cells.

\begin{figure*}
	\includegraphics[width=\linewidth]{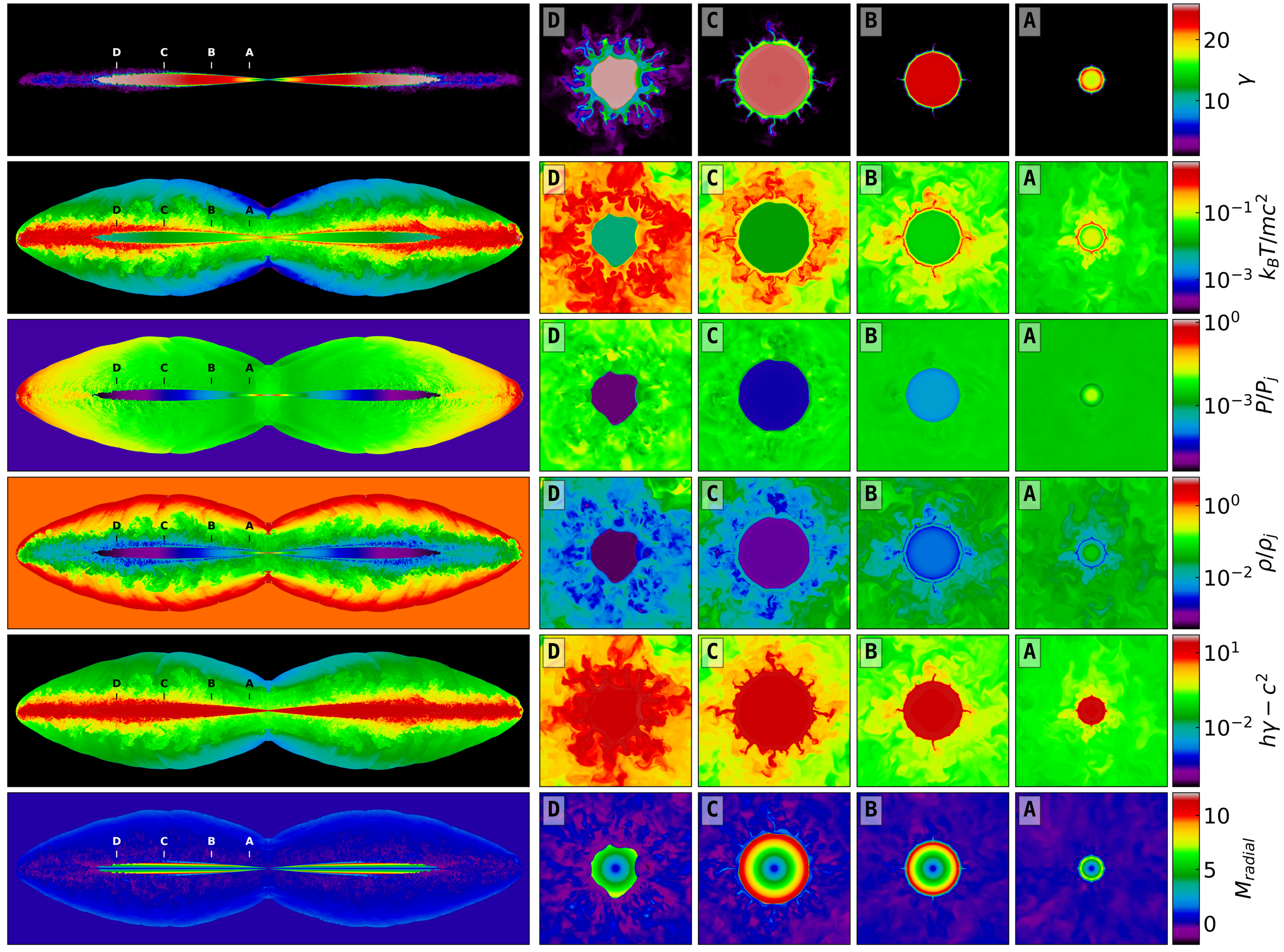}
    \caption{
    From top to bottom: Lorentz factor, temperature, pressure (normalized by the jet source pressure $p_{\text{j}}$), proper mass density (normalized by the jet source density $\rho_{\text{j}}$), relativistic Bernoulli number minus $c^2$ (i.e. $h\gamma-c^2$), and radial component of Mach number defined by \Cref{eq:transverse Mach number} at $t=0. 73 L/c$. Left column: longitudinal slices passing through the jet source. Right four columns: transverse slices passing through the locations labelled by A, B, C, and D.}
   \label{fig:Limb_brightened_jet}
\end{figure*}

\Cref{fig:Limb_brightened_jet} shows the simulation results. It demonstrates that the jet flow is entirely confined by a turbulent cocoon at all time. Two points are worth noting from these longitudinal slices. First, the Lorentz factor (first row) rises from $10$ to $26$, and meanwhile the temperature (second row) drops from 0.5 to 0.01 along the jet. Second, the relativistic Bernoulli number minus $c^2$ (fifth row), defined as $h\gamma-c^2$, remains nearly constant within the spine region. According to the de Laval nozzle effect, these suggest that thermal energy is converted to kinetic energy by the expansion of a supersonic flow. Surprisingly, the gases are still accelerated in the region between the label `C' and the confinement point close to `D'. These images seem to suggest acceleration during flow convergence, which in fact does not contradict the de Laval nozzle effect. The gases still expands away from the jet axis after passing `C', which can be confirmed by examining the transverse slice of the radial component of the Mach number (the last row),
\begin{equation}
     \mathscr{M}_{\text{radial}}=\left(\hat{r}\cdot \mathbf{U}\right)/\left(c_{s}\gamma_{s}\right),
     \label{eq:transverse Mach number}
 \end{equation}
where $\hat{r}$ is the cylindrical unit radial vector, $\mathbf{U}$ is the four-velocity of flow, and $\gamma_{s}=c_{s}/\sqrt{1-c_{s}^2}$. The definition of the radial Mach number given by \Cref{eq:transverse Mach number} is Lorentz invariant when the transforming direction is along the jet. Obviously, gases expand not only between the jet source and `C', but also inside the entire central spine region. Thus, the flow convergence in between `C' and `D' is a false impression.

Associated with this expanding jet flow is the limb-brightened phenomenon. Confined by the cocoon, the radial flow of the cooler jet imparts onto the cocoon with a boundary shock, as signified by the edge $\mathscr{M}_{\text{radial}} \gg 1$. Hot gases in the post-shock region then diffuse into the cocoon transverse to the jet through some instabilities composed of high-density and low-temperature fingers. This finger pattern is similar to that reported in Section \ref{Multi-dimensional grid effects}.

Certainly the boundary shock can generate particle acceleration and produce extra synchrotron brightness at the jet edge, thus yielding limb brightening. Since the boundary shock is relatively weak, the extra synchrotron brightness cannot be immense. This may explain why limb brightening is mostly observed in nearby AGN jets.

\section{conclusions}
\label{conclusions}
 In this paper, we have presented a novel special relativistic hydrodynamics code, \textsc{gamer-sr}, which incorporates a new, well-tailored conversion scheme (cf. \Cref{fig:flowchart}) between primitive and conserved variables, together with the Taub-Mathews equation of state (TM EoS) covering both the ultra-relativistic and non-relativistic limits. The new scheme adopts the four-velocity $U^{j}$, the reduced energy density $\tilde{E}$, and the reduced enthalpy $\tilde{h}$ to effectively avoid the catastrophic cancellation in subsonic flows at all finite temperature, including the particularly challenging low-temperature regime, with errors decreasing as $\mathscr{M}^2\epsilon_{\text{machine}}$ when $\mathscr{M}\gg1$.
 
 We have numerically derived the exact solution of a Riemann problem covering both extreme cold and ultra-relativistically hot gases with the TM EoS. Simulation results using our new scheme are in very good agreement with the exact solution in both the ultra-relativistic and non-relativistic regimes. (cf. \Cref{fig:non-relativistic shock tube}). In comparison, the catastrophic cancellation arising from the original (unoptimized) scheme can be much more severe than the truncation error in the non-relativistic limit, especially in the region swept by a traveling contact discontinuity.

The new scheme has been integrated into the code \textsc{gamer} to facilitate the hybrid OpenMP/MPI/GPU parallelization and adaptive mesh refinement. Thanks to that, the performance of the root-finding iterations in the TM EoS can be significantly improved by GPU. The parallel efficiency using 2048 computing nodes is measured to be $\sim$ 45 per cent for strong scaling (cf. \Cref{fig:strong scaling}) and $\sim$ 75 per cent for weak scaling (cf. \Cref{fig:weak scaling}) on the \texttt{Piz-Daint} supercomputer.

\textsc{gamer-sr} has been demonstrated to be able to handle ultra-relativistic flow with a Lorentz factor as high as $10^6$. However, we also find that the Cartesian grids can lead to artificial dissipation when the direction of a high Mach number flow is not aligned with grids. This problem cannot be mitigated by increasing spatial and temporal resolution.
 
 We have examined two astrophysical problems with coexisting relativistically hot and cold gases to demonstrate the power of \textsc{gamer-sr}. The first problem deals with a relativistic blast wave with a triaxial source. Not only do we find that the code is able to capture the ultra-relativistic strong shock very well, but we also discover a simple rule governing how the triaxiality of the blast wave diminishes as a function of the blast wave radius.

 The second problem addresses the flow acceleration and limb-brightening of a relativistic AGN jet. We find that the jet, from its head to its source, is always enclosed inside a turbulent cocoon. The jet is accelerated all the way up to the first confinement point, where an internal shock appears. We attribute such flow acceleration to the relativistic Bernoulli's law. In addition, the synchrotron limb-brightening is found to be caused by the jet transverse boundary shock, outside which the post-shock cosmic-ray particles are mixed with the turbulent cocoon and give out extra synchrotron emission.

\section*{Acknowledgement}
P.T. thanks Kuo-Chuan Pan for helping conduct the parallel scaling tests at CSCS ($\texttt{Piz-Daint}$) under Grant No. 661.  H.S. acknowledges funding support from the Jade Mountain Young Scholar Award No. NTU-109V0201, sponsored by the Ministry of Education, Taiwan. This research is partially supported by the Ministry of Science and Technology (MOST) of Taiwan under Grants No. MOST 107-2119-M-002-036-MY3 and No. MOST 108-2112-M-002-023-MY3, and the NTU Core Consortium project under Grants No. NTU-CC-108L893401 and No. NTU-CC-108L893402.

\section*{Data Availability}
The data underlying this article are available in the article and in its online supplementary material.




\bibliographystyle{mnras}
\bibliography{v1} 

\begin{thebibliography}{}
\makeatletter
\relax
\def\mn@urlcharsother{\let\do\@makeother \do\$\do\&\do\#\do\^\do\_\do\%\do\~}
\def\mn@doi{\begingroup\mn@urlcharsother \@ifnextchar [ {\mn@doi@}
  {\mn@doi@[]}}
\def\mn@doi@[#1]#2{\def\@tempa{#1}\ifx\@tempa\@empty \href
  {http://dx.doi.org/#2} {doi:#2}\else \href {http://dx.doi.org/#2} {#1}\fi
  \endgroup}
\def\mn@eprint#1#2{\mn@eprint@#1:#2::\@nil}
\def\mn@eprint@arXiv#1{\href {http://arxiv.org/abs/#1} {{\tt arXiv:#1}}}
\def\mn@eprint@dblp#1{\href {http://dblp.uni-trier.de/rec/bibtex/#1.xml}
  {dblp:#1}}
\def\mn@eprint@#1:#2:#3:#4\@nil{\def\@tempa {#1}\def\@tempb {#2}\def\@tempc
  {#3}\ifx \@tempc \@empty \let \@tempc \@tempb \let \@tempb \@tempa \fi \ifx
  \@tempb \@empty \def\@tempb {arXiv}\fi \@ifundefined
  {mn@eprint@\@tempb}{\@tempb:\@tempc}{\expandafter \expandafter \csname
  mn@eprint@\@tempb\endcsname \expandafter{\@tempc}}}

\bibitem[\protect\citeauthoryear{Asada \& Nakamura}{Asada \&
  Nakamura}{2012}]{Asada2012}
Asada K.,  Nakamura M.,  2012, \mn@doi [The Astrophysical Journal]
  {10.1088/2041-8205/745/2/l28}, 745, L28

\bibitem[\protect\citeauthoryear{Blandford, Meier  \& Readhead}{Blandford
  et~al.}{2019}]{Blandford2018}
Blandford R.,  Meier D.,   Readhead A.,  2019, \mn@doi [Annual Review of
  Astronomy and Astrophysics] {10.1146/annurev-astro-081817-051948}, 57, 467

\bibitem[\protect\citeauthoryear{Boccardi, Krichbaum, Bach, Mertens, Ros, Alef
  \& Zensus}{Boccardi et~al.}{2015}]{Boccardi2015}
Boccardi B.,  Krichbaum T.~P.,  Bach U.,  Mertens F.,  Ros E.,  Alef W.,
  Zensus J.~A.,  2015, \mn@doi [A and A] {10.1051/0004-6361/201526985}, 585,
  A33

\bibitem[\protect\citeauthoryear{Chiueh}{Chiueh}{1989}]{Chiue-PhysRevLett.63.113}
Chiueh T.,  1989, \mn@doi [Phys. Rev. Lett.] {10.1103/PhysRevLett.63.113}, 63,
  113

\bibitem[\protect\citeauthoryear{{Chiueh}, {Li}  \& {Begelman}}{{Chiueh}
  et~al.}{1991}]{Chiueh-1991ApJ...377..462C}
{Chiueh} T.,  {Li} Z.-Y.,   {Begelman} M.~C.,  1991, \mn@doi [\apj]
  {10.1086/170375}, \href
  {https://ui.adsabs.harvard.edu/abs/1991ApJ...377..462C} {377, 462}

\bibitem[\protect\citeauthoryear{Chiueh, Li  \& Begelman}{Chiueh
  et~al.}{1998}]{Chiueh_1998}
Chiueh T.,  Li Z.-Y.,   Begelman M.~C.,  1998, \mn@doi [The Astrophysical
  Journal] {10.1086/306209}, 505, 835

\bibitem[\protect\citeauthoryear{Falle}{Falle}{1991}]{VL1}
Falle S. A. E.~G.,  1991, \mn@doi [Monthly Notices of the Royal Astronomical
  Society] {10.1093/mnras/250.3.581}, 250, 581

\bibitem[\protect\citeauthoryear{{Fong} et~al.,}{{Fong} et~al.}{2019}]{NM4}
{Fong} W.,  et~al., 2019, \mn@doi [\apjl] {10.3847/2041-8213/ab3d9e}, \href
  {https://ui.adsabs.harvard.edu/abs/2019ApJ...883L...1F} {883, L1}

\bibitem[\protect\citeauthoryear{Fryxell et~al.,}{Fryxell et~al.}{2000}]{FLASH}
Fryxell B.,  et~al., 2000, \mn@doi [The Astrophysical Journal Supplement
  Series] {10.1086/317361}, 131, 273

\bibitem[\protect\citeauthoryear{{Ghirlanda} et~al.,}{{Ghirlanda}
  et~al.}{2019}]{NM3}
{Ghirlanda} G.,  et~al., 2019, \mn@doi [Science] {10.1126/science.aau8815},
  \href {https://ui.adsabs.harvard.edu/abs/2019Sci...363..968G} {363, 968}

\bibitem[\protect\citeauthoryear{Giovannini et~al.,}{Giovannini
  et~al.}{2018}]{Giovannini2018}
Giovannini G.,  et~al., 2018, Nature Astronomy, 2, 472

\bibitem[\protect\citeauthoryear{Giroletti et~al.,}{Giroletti
  et~al.}{2004}]{Giroletti2004}
Giroletti M.,  et~al., 2004, \mn@doi [The Astrophysical Journal]
  {10.1086/379663}, 600, 127

\bibitem[\protect\citeauthoryear{Gourgouliatos \& Komissarov}{Gourgouliatos \&
  Komissarov}{2017}]{Gourgouliatos2017}
Gourgouliatos K.~N.,  Komissarov S.~S.,  2017, \mn@doi [Nature Astronomy]
  {10.1038/s41550-017-0338-3}, 2, 167

\bibitem[\protect\citeauthoryear{Higham}{Higham}{2002}]{Nicholas}
Higham N.~J.,  2002, Accuracy and Stability of Numerical Algorithms, 2nd edn.
Society for Industrial and Applied Mathematics, USA

\bibitem[\protect\citeauthoryear{Jüttner}{Jüttner}{1911}]{Juttner}
Jüttner F.,  1911, \mn@doi [Annalen der Physik] {10.1002/andp.19113390503},
  339, 856

\bibitem[\protect\citeauthoryear{{Kennel} \& {Coroniti}}{{Kennel} \&
  {Coroniti}}{1984a}]{1984ApJ...283..694K}
{Kennel} C.~F.,  {Coroniti} F.~V.,  1984a, \mn@doi [\apj] {10.1086/162356},
  \href {https://ui.adsabs.harvard.edu/abs/1984ApJ...283..694K} {283, 694}

\bibitem[\protect\citeauthoryear{{Kennel} \& {Coroniti}}{{Kennel} \&
  {Coroniti}}{1984b}]{1984ApJ...283..710K}
{Kennel} C.~F.,  {Coroniti} F.~V.,  1984b, \mn@doi [\apj] {10.1086/162357},
  \href {https://ui.adsabs.harvard.edu/abs/1984ApJ...283..710K} {283, 710}

\bibitem[\protect\citeauthoryear{Kim et~al.,}{Kim et~al.}{2018}]{Kim2018}
Kim J.-Y.,  et~al., 2018, A\&A, 616

\bibitem[\protect\citeauthoryear{{Li}, {Chiueh}  \& {Begelman}}{{Li}
  et~al.}{1992}]{Chiueh-1992ApJ...394..459L}
{Li} Z.-Y.,  {Chiueh} T.,   {Begelman} M.~C.,  1992, \mn@doi [\apj]
  {10.1086/171597}, \href
  {https://ui.adsabs.harvard.edu/abs/1992ApJ...394..459L} {394, 459}

\bibitem[\protect\citeauthoryear{Lora-Clavijo, Cruz-Osorio  \&
  Guzm{\'{a}}n}{Lora-Clavijo et~al.}{2015}]{CAFE}
Lora-Clavijo F.~D.,  Cruz-Osorio A.,   Guzm{\'{a}}n F.~S.,  2015, \mn@doi [The
  Astrophysical Journal Supplement Series] {10.1088/0067-0049/218/2/24}, 218,
  24

\bibitem[\protect\citeauthoryear{Martí \& Müller}{Martí \&
  Müller}{1994}]{Marti1994}
Martí J.~M.,  Müller E.,  1994, \mn@doi [Journal of Fluid Mechanics]
  {10.1017/s0022112094003344}, 258, 317

\bibitem[\protect\citeauthoryear{{Mathews}}{{Mathews}}{1971}]{TM_EOS}
{Mathews} W.~G.,  1971, \mn@doi [\apj] {10.1086/150883}, \href
  {https://ui.adsabs.harvard.edu/abs/1971ApJ...165..147M} {165, 147}

\bibitem[\protect\citeauthoryear{Mignone \& Bodo}{Mignone \&
  Bodo}{2005}]{HLLC_srhydro}
Mignone A.,  Bodo G.,  2005, \mn@doi [Monthly Notices of the Royal Astronomical
  Society] {10.1111/j.1365-2966.2005.09546.x}, 364, 126

\bibitem[\protect\citeauthoryear{Mignone \& Bodo}{Mignone \&
  Bodo}{2006}]{HLLC_srmhd}
Mignone A.,  Bodo G.,  2006, \mn@doi [Monthly Notices of the Royal Astronomical
  Society] {10.1111/j.1365-2966.2006.10162.x}, 368, 1040

\bibitem[\protect\citeauthoryear{Mignone \& McKinney}{Mignone \&
  McKinney}{2007}]{NR_Limit}
Mignone A.,  McKinney J.~C.,  2007, \mn@doi [Monthly Notices of the Royal
  Astronomical Society] {10.1111/j.1365-2966.2007.11849.x}, 378, 1118

\bibitem[\protect\citeauthoryear{Mignone, Plewa  \& Bodo}{Mignone
  et~al.}{2005}]{Compare_TM_EOS}
Mignone A.,  Plewa T.,   Bodo G.,  2005, \mn@doi [The Astrophysical Journal
  Supplement Series] {10.1086/430905}, 160

\bibitem[\protect\citeauthoryear{{Mooley} et~al.,}{{Mooley}
  et~al.}{2018a}]{NM1}
{Mooley} K.~P.,  et~al., 2018a, \mn@doi [\nat] {10.1038/nature25452}, \href
  {https://ui.adsabs.harvard.edu/abs/2018Natur.554..207M} {554, 207}

\bibitem[\protect\citeauthoryear{{Mooley} et~al.,}{{Mooley}
  et~al.}{2018b}]{NM2}
{Mooley} K.~P.,  et~al., 2018b, \mn@doi [\nat] {10.1038/s41586-018-0486-3},
  \href {https://ui.adsabs.harvard.edu/abs/2018Natur.561..355M} {561, 355}

\bibitem[\protect\citeauthoryear{Nagai et~al.,}{Nagai
  et~al.}{2014}]{Nagai_2014}
Nagai H.,  et~al., 2014, \mn@doi [The Astrophysical Journal]
  {10.1088/0004-637x/785/1/53}, 785, 53

\bibitem[\protect\citeauthoryear{Noble, Gammie, McKinney  \& Zanna}{Noble
  et~al.}{2006}]{Noble_2006}
Noble S.~C.,  Gammie C.~F.,  McKinney J.~C.,   Zanna L.~D.,  2006, \mn@doi [The
  Astrophysical Journal] {10.1086/500349}, 641, 626

\bibitem[\protect\citeauthoryear{{N{\'u}{\~n}ez-de la Rosa} \&
  {Munz}}{{N{\'u}{\~n}ez-de la Rosa} \& {Munz}}{2016}]{XTROEM}
{N{\'u}{\~n}ez-de la Rosa} J.,  {Munz} C.-D.,  2016, \mn@doi [\mnras]
  {10.1093/mnras/stw999}, \href
  {https://ui.adsabs.harvard.edu/abs/2016MNRAS.460..535N} {460, 535}

\bibitem[\protect\citeauthoryear{Rezzolla \& Zanotti}{Rezzolla \&
  Zanotti}{2018}]{Rezzolla2018}
Rezzolla L.,  Zanotti O.,  2018, Relativistic hydrodynamics.
Oxford University Press

\bibitem[\protect\citeauthoryear{Rezzolla, Zanotti  \& Pons}{Rezzolla
  et~al.}{2001}]{REZZOLLA2001}
Rezzolla L.,  Zanotti O.,   Pons J.~A.,  2001, \mn@doi [Journal of Fluid
  Mechanics] {10.1017/s0022112001006450}, 449, 395

\bibitem[\protect\citeauthoryear{Ryu, Chattopadhyay  \& Choi}{Ryu
  et~al.}{2006}]{RC_EOS}
Ryu D.,  Chattopadhyay I.,   Choi E.,  2006, \mn@doi [The Astrophysical Journal
  Supplement Series] {10.1086/505937}, 166, 410

\bibitem[\protect\citeauthoryear{Schive, Tsai  \& Chiueh}{Schive
  et~al.}{2010}]{gamer-1}
Schive H.-Y.,  Tsai Y.-C.,   Chiueh T.,  2010, \mn@doi [The Astrophysical
  Journal Supplement Series] {10.1088/0067-0049/186/2/457}, 186, 457

\bibitem[\protect\citeauthoryear{Schive, ZuHone, Goldbaum, Turk, Gaspari  \&
  Cheng}{Schive et~al.}{2018}]{gamer-2}
Schive H.-Y.,  ZuHone J.~A.,  Goldbaum N.~J.,  Turk M.~J.,  Gaspari M.,   Cheng
  C.-Y.,  2018, \mn@doi [Monthly Notices of the Royal Astronomical Society]
  {10.1093/mnras/sty2586}, 481, 4815

\bibitem[\protect\citeauthoryear{Sod}{Sod}{1978}]{SOD1978}
Sod G.~A.,  1978, \mn@doi [Journal of Computational Physics]
  {https://doi.org/10.1016/0021-9991(78)90023-2}, 27, 1

\bibitem[\protect\citeauthoryear{Synge}{Synge}{1957}]{Synge}
Synge J.~L.,  1957, North-Holland Pub. Co.; Interscience Publishers

\bibitem[\protect\citeauthoryear{{Taub}}{{Taub}}{1948}]{Taub}
{Taub} A.~H.,  1948, \mn@doi [Physical Review] {10.1103/PhysRev.74.328}, \href
  {https://ui.adsabs.harvard.edu/abs/1948PhRv...74..328T} {74, 328}

\bibitem[\protect\citeauthoryear{Toro}{Toro}{2011}]{Toro}
Toro E.~F.,  2011, Riemann solvers and numerical methods for fluid dynamics : a
  practical introduction.
Springer, Berlin

\bibitem[\protect\citeauthoryear{{Turk}, {Smith}, {Oishi}, {Skory}, {Skillman},
  {Abel}  \& {Norman}}{{Turk} et~al.}{2011}]{YT}
{Turk} M.~J.,  {Smith} B.~D.,  {Oishi} J.~S.,  {Skory} S.,  {Skillman} S.~W.,
  {Abel} T.,   {Norman} M.~L.,  2011, \mn@doi [The Astrophysical Journal
  Supplement Series] {10.1088/0067-0049/192/1/9}, \href
  {https://ui.adsabs.harvard.edu/abs/2011ApJS..192....9T} {192, 9}

\bibitem[\protect\citeauthoryear{{Woosley}}{{Woosley}}{1993}]{LongGRB}
{Woosley} S.~E.,  1993, \mn@doi [\apj] {10.1086/172359}, \href
  {https://ui.adsabs.harvard.edu/abs/1993ApJ...405..273W} {405, 273}

\bibitem[\protect\citeauthoryear{van Leer}{van Leer}{1979}]{van_Leer_1979}
van Leer B.,  1979, \mn@doi [Journal of Computational Physics]
  {10.1016/0021-9991(79)90145-1}, 32, 101

\bibitem[\protect\citeauthoryear{van Leer}{van Leer}{2006}]{VL2}
van Leer B.,  2006, pp 192--206

\makeatother
\end{thebibliography}




\appendix
Note that the speed of light, the particle mass, and the Boltzmann constant are set to unity in Appendix for simplicity.
\definecolor{AppendixAf}{RGB}{186, 158, 54}
\definecolor{AppendixAfHot}{RGB}{214, 49, 34}
\definecolor{AppendixAfCold}{RGB}{103, 138, 235}
\definecolor{AppendixAfHori}{RGB}{57, 204, 113}

\section{Initial guess for Newton-Raphson iteration}
\label{The choice of initial guesses}
We use the Newton-Raphson iteration to find the root $\tilde{h}$ of Equation~(\ref{transcendental equation}). The iteration requires the derivative of \Cref{transcendental equation} with respect to $\tilde{h}$:
\begin{equation}
\frac{df}{d\tilde{h}}=2\tilde{h}+2-2T-2\left(\tilde{h}+1\right)\frac{dT}{d\tilde{h}}+\frac{2T\left(\tilde{h}+1\right)^4\frac{dT}{d\tilde{h}}+2T\left(\tilde{h}+1\right)\left(\frac{\abs{\mathbf{M}}}{D}\right)^2\left[\left(\tilde{h}+1\right)\frac{dT}{d\tilde{h}}+T\right]}{\left[\left(\tilde{h}+1\right)^2+\left(\frac{\abs{\mathbf{M}}}{D}\right)^2\right]^2},
\end{equation}
where
\begin{equation}
\frac{dT}{d\tilde{h}}=\frac{4\tilde{h}+4}{5\tilde{h}+5+\sqrt{9\tilde{h}^2+18\tilde{h}+25}}
-\frac{\left(2\tilde{h}^2+4\tilde{h}\right)\left(\frac{9\tilde{h}+9}{\sqrt{9\tilde{h}^2+18\tilde{h}+25}}+5\right)}{\left(5\tilde{h}+5+\sqrt{9\tilde{h}^2+18\tilde{h}+25}\right)^2},
\label{dT_dh}
\end{equation}
follows from \Cref{T of h}.

\begin{figure}
\includegraphics[scale=0.7]{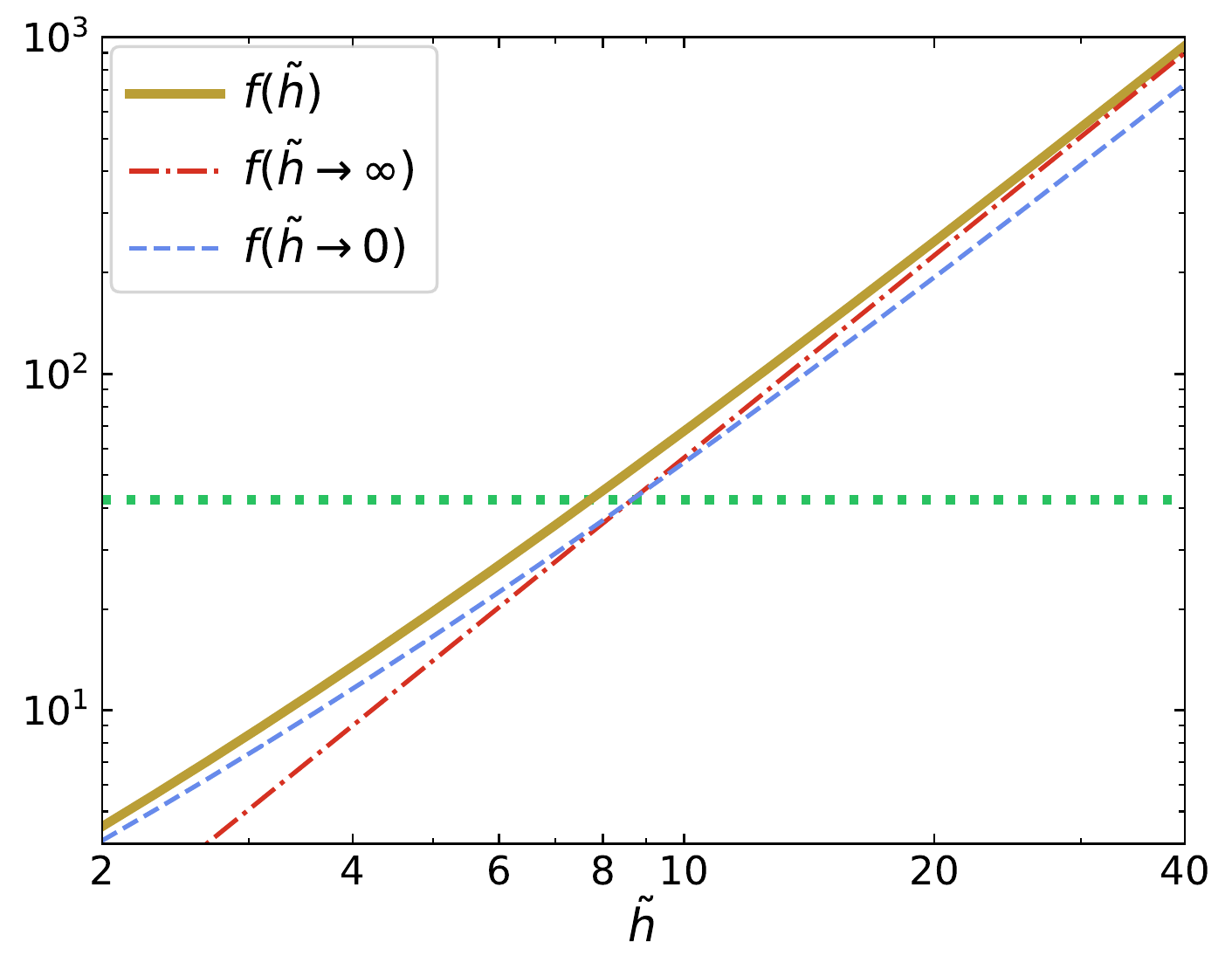}
\caption{$f(\tilde{h};\abs{\mathbf{M}}/D=1)$ ($\MySolidLine[draw=AppendixAf,fill=AppendixAf]$) and its asymptotes when $\tilde{h}\rightarrow \infty$ ($\MyDashedDottedLine[draw=AppendixAfHot,fill=AppendixAfHot]$) and $\tilde{h}\rightarrow 0$ ($\MyDashedLine[draw=AppendixAfCold,fill=AppendixAfCold]$). The horizontal line ($\MyDottedLine[draw=AppendixAfHori,fill=AppendixAfHori]$) is given by Equation \ref{ColdHot}, which passes through the intersection of two asymptotes and provides a threshold to distinguish between `cold' and `hot' gases for the initial guess of $\tilde{h}$ in the Newton-Raphson iteration.}
\label{fig:FunHTilde}
\end{figure}

The root-finding iteration also requires an initial guess of $\tilde{h}_{\text{guess}}$, for which we suggest the following procedure. In the low-$T$ limit, we Taylor expand Equation (\ref{transcendental equation}) in powers of $\tilde{h}$ and keep the first- and second-order terms:
\begin{equation}
\left(\frac{\tilde{E}}{D}\right)^2+2\left(\frac{\tilde{E}}{D}\right)-\left(\frac{\abs{\mathbf{M}}}{D}\right)^2
=\frac{6}{5}\tilde{h}+\left\{\frac{43}{125}+\frac{4}{25\left[1+\left(\frac{\abs{\mathbf{M}}}{D}\right)^2\right]}\right\}\tilde{h}^2.
\label{cold limit of transcendental eq}
\end{equation}
Solving Equation~(\ref{cold limit of transcendental eq}) for the unknown $\tilde{h}$ gives the positive solution:
\begin{equation}
\tilde{h}_{\text{guess}}
=\frac{
\sqrt{125\left[1+\left(\frac{\abs{\mathbf{M}}}{D}\right)^2\right]\left[43\left(\frac{\abs{\mathbf{M}}}{D}\right)^2+63\right]\left[\left(\frac{\tilde{E}}{D}\right)^2+2\left(\frac{\tilde{E}}{D}\right)-\left(\frac{\abs{\mathbf{M}}}{D}\right)^2\right]}}
{\left[43\left(\frac{\abs{\mathbf{M}}}{D}\right)^2+63\right]\left\{75\left[1+\left(\frac{\abs{\mathbf{M}}}{D}\right)^2\right]+\sqrt{125\left[1+\left(\frac{\abs{\mathbf{M}}}{D}\right)^2\right]\left[43\left(\frac{\abs{\mathbf{M}}}{D}\right)^2+63\right]\left[\left(\frac{\tilde{E}}{D}\right)^2+2\left(\frac{\tilde{E}}{D}\right)-\left(\frac{\abs{\mathbf{M}}}{D}\right)^2\right]+75^2\left[1+\left(\frac{\abs{\mathbf{M}}}{D}\right)^2\right]^2}\right\}}.
\label{guess for cold fluid}
\end{equation} In the opposite high-$T$ limit, Equation~(\ref{transcendental equation}) can be reduced to
\begin{equation}
\left(\frac{\tilde{E}}{D}\right)^2+2\left(\frac{\tilde{E}}{D}\right)-\left(\frac{\abs{\mathbf{M}}}{D}\right)^2
=\frac{9}{16}\tilde{h}^2,
\label{hot limit of transcendental eq}
\end{equation}
which leads to
\begin{equation}
\tilde{h}_{\text{guess}}=
\frac{4}{3}\sqrt{\left(\frac{\tilde{E}}{D}\right)^2+2\left(\frac{\tilde{E}}{D}\right)-\left(\frac{\abs{\mathbf{M}}}{D}\right)^2}.
\label{guess for hot fluid}
\end{equation}
Equation~(\ref{guess for cold fluid}) and Equation~(\ref{guess for hot fluid}) provide two initial guesses for `cold' and `hot' gases, respectively. The threshold to distinguish between `cold' gases and `hot' gases is given by
\begin{equation}
\left\{\frac{1800\left[1+\left(\frac{\abs{\mathbf{M}}}{D}\right)^2\right]}{437\left(\frac{\abs{\mathbf{M}}}{D}\right)^2+117}\right\}^2,
\label{ColdHot}
\end{equation}
which is obtained by equating Equations (\ref{guess for cold fluid}) and (\ref{guess for hot fluid}) (see \Cref{fig:FunHTilde}). If $(\tilde{E}/D)^2+2(\tilde{E}/D)-\left(\abs{\mathbf{M}}/D\right)^2$ is greater than \Cref{ColdHot}, we choose Equation~(\ref{guess for hot fluid}) as an initial guess for the Newton-Raphson iteration (hot gases); otherwise, we choose \Cref{guess for cold fluid} (cold gases).

\begin{figure}
\includegraphics[width=\linewidth]{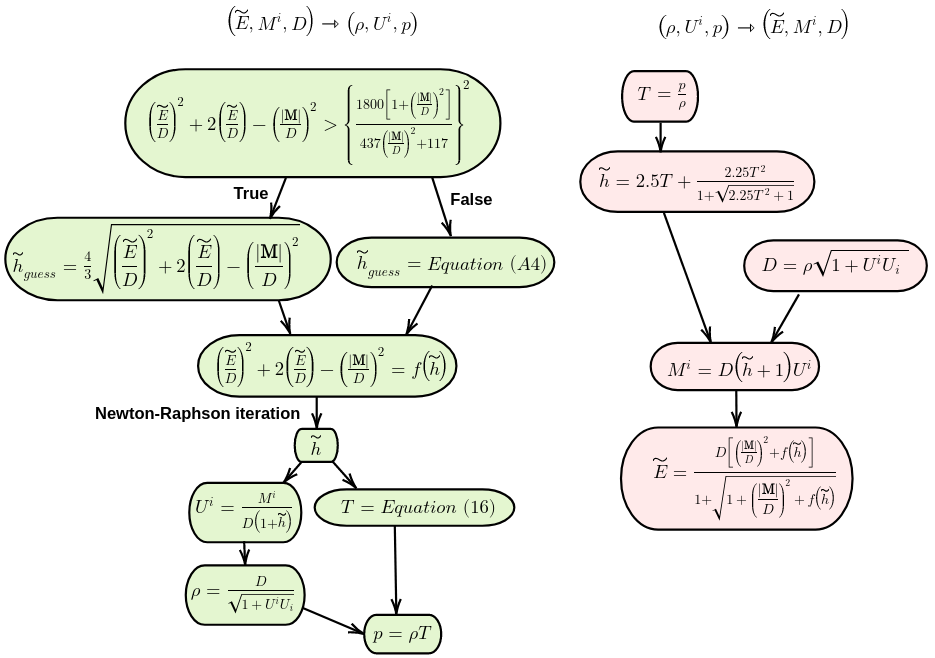}
\caption{Flowchart of converting conserved variables to primitive variables (left) and the opposite (right).}
\label{fig:flowchart}
\end{figure}

\definecolor{ErrorURNumerical}{RGB}{1,129,74}
\definecolor{ErrorURExact}{RGB}{224, 201, 73}
\definecolor{ErrorNRNumerical}{RGB}{255,0,0}
\definecolor{ErrorNRExact}{RGB}{0,0,255}
\section{Numerical error analysis for root-finding}
\label{Appendix:Numerical error analysis}
\begin{figure}
\includegraphics[scale=0.8]{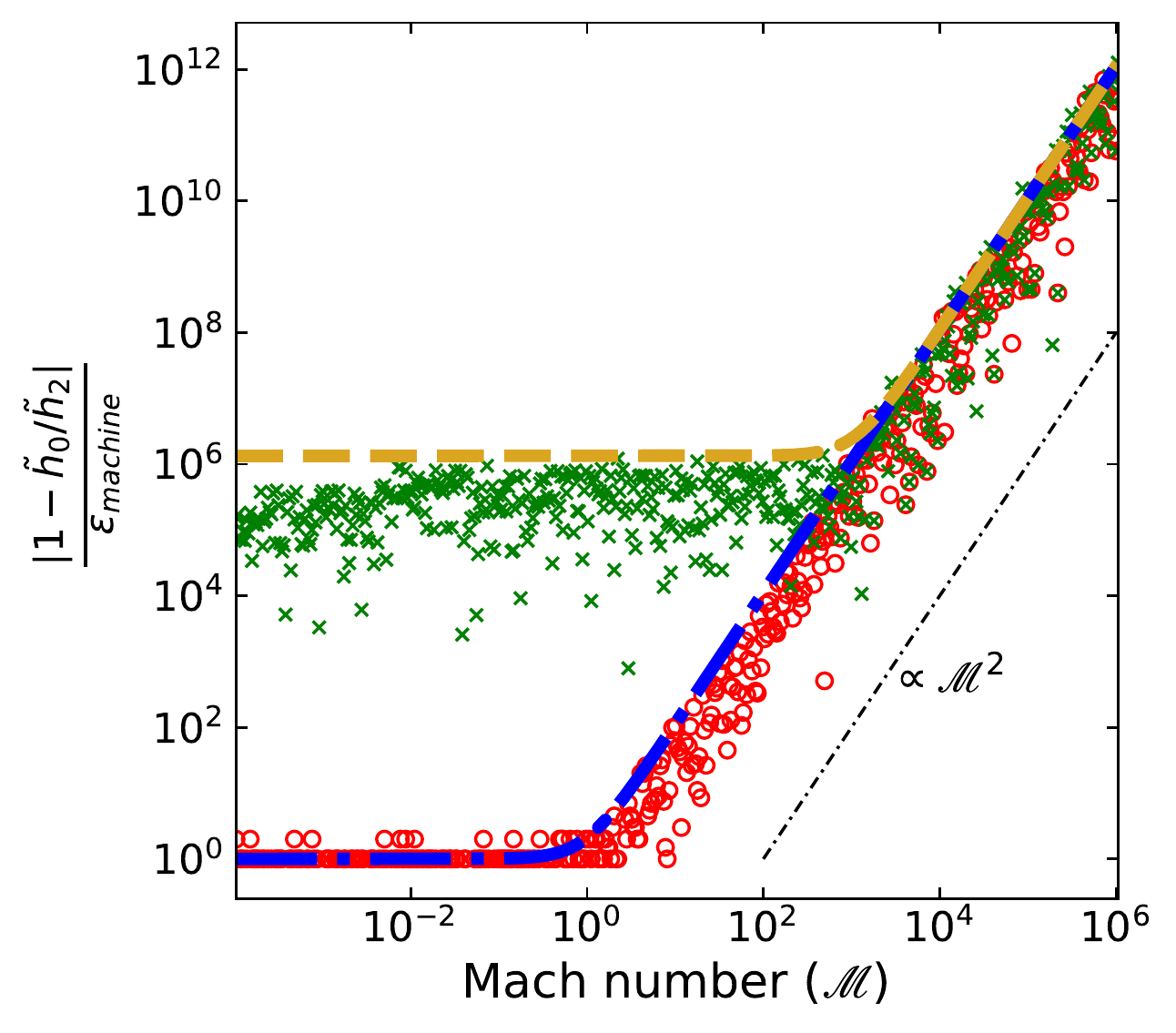}
\caption{Numerical errors of the conversion between primitive and conserved variables. It shows the relative error $\abs{1-\tilde{h}_{0}/\tilde{h}_{2}}$ as a function of Mach number with a given non-relativistic temperature ($k_{B}T/mc^2=10^{-8}$) for the new scheme ($\tikz\draw[ErrorNRNumerical,fill=white,line width=1] (0,0) circle (.5ex);$) and the original scheme ($\MyCross[draw=ErrorURNumerical,fill=ErrorURNumerical]$). These errors are mainly caused by the cancellation in $\left(\tilde{E}/D\right)^2+2\left(\tilde{E}/D\right)-\left(\abs{\mathbf{M}}/D\right)^2$ and $\left(E/D\right)^2-\left(\abs{\mathbf{M}}/D\right)^2-1$, which are inevitably introduced during the root-finding iteration and consistent with the predicted values given by \Cref{eq:ImprovedRelativeError} ($\MyDashedDottedLine[ErrorNRExact,fill=white]$) and \Cref{eq:OriRelativeError}                   ($\MyDottedLine[draw=ErrorURExact,fill=ErrorURExact]$).}
\label{fig:ErrorAnalysis}
\end{figure}

\Cref{fig:flowchart} provides a detailed flowchart of the conversion between primitive and conserved variables. \Cref{fig:ErrorAnalysis} demonstrates that the numerical errors of root-finding arising from the new and original conversion schemes are consistent with the predicted values given by \Cref{eq:ImprovedRelativeError} and \Cref{eq:OriRelativeError}. We measure this conversion error by first converting the input primitive variables $(\rho_0, U^{i}_0, p_0)$ into conserved variables $(D_1, M^{i}_1, \tilde{E}_1)$. Next, we convert $(D_1, M^{i}_1, \tilde{E}_1)$ back to $(\rho_2, U^{i}_2, p_2)$ and then measure the relative error between $\tilde{h}(p_0/\rho_0)\coloneqq\tilde{h}_{0}$ and $\tilde{h}(p_2/\rho_2)\coloneqq\tilde{h}_{2}$. Since the catastrophic cancellation is more prominent in the low-$T$ limit, we measure the error $\abs{1-\tilde{h}_{0}/\tilde{h}_{2}}$ as a function of Mach number from $10^{-4}$ to $10^{6}$ with a fixed non-relativistic temperature $k_{B}T/mc^2=10^{-8}$ (i.e. the blue dashed-dotted line in \Cref{fig:ErrorDistribution}). To verify the accuracy in three-dimensional space, we choose the direction of four-velocity to be parallel to the line $x=y=z$ (i.e. $U^i_0=\abs{\mathbf{U}_0}/\sqrt{3}$ for all $i$). Double precision is adopted to handle the large dynamic range. \Cref{fig:ErrorAnalysis} confirms that the numerical errors of the new and original schemes are mainly caused by round-off errors in the calculation of $\left(\tilde{E}/D\right)^2+2\left(\tilde{E}/D\right)-\left(\abs{\mathbf{M}}/D\right)^2$ and $\left(E/D\right)^2-\left(\abs{\mathbf{M}}/D\right)^2-1$, respectively.

\section{Exact solutions of relativistic Riemann problems with the TM equation of state}
\label{appendix:exact solution}
To derive the exact solutions of relativistic Riemann problems with the TM EoS, we have generalized the previous framework of a constant polytropic EoS (\citealt{Marti1994}; \citealt{REZZOLLA2001}) to the TM EoS. More precisely, this approach can be applied to any EoS once we know the relationship between enthalpy and temperature. Here we only summarize the important equations and highlight salient differences from the polytropic EoS. We use the subscripts $\text{$L/C_{L}/C_{R}/R$}$ to refer to the left/left-contact/right-contact/right regions and define the relative four-velocity of $U_{1}$ with respect to $U_{2}$ as $\texttt{Relative}(U_{1}, U_{2})$. Note that we have replaced three-velocity with four-velocity again to avoid catastrophic cancellation in the ultra-relativistic limit.

The exact solution of a relativistic Riemann problem with the TM EoS can be obtained through the following three steps:
\begin{enumerate}
    \item For a given initial condition, we can determine the wave pattern by comparing the relative velocity between the two unperturbed initial states with the three limiting values. These values mark the transition from one wave pattern to another and can be directly computed from the initial condition. See \citet{REZZOLLA2001} for details.
    \item We determine the unknown pressure $p^{*}$ between the left and right waves by numerically solving
    \begin{equation}
    U_{LR}=\texttt{Relative}(U_{LC_{L}}(p^{*}), U_{RC_{R}}(p^{*})),
    \label{eq:p*}
    \end{equation}
    where $U_{LR}=\texttt{Relative}(U_{L}, U_{R})$, $U_{LC_{L}}(p^{*})=\texttt{Relative}(U_{L}, U_{C_{L}}(p^{*}))$, and $U_{RC_{R}}(p^{*})=\texttt{Relative}(U_{R}, U_{C_{R}}(p^{*}))$. Note that the four-velocity in the left-/right-contact region, $U_{C_{L/R}}$, is different for each of the three possible wave patterns. For example, if the left wave is rarefaction and the right wave is shock, then $U_{C_{L}}=U_{\mathscr{R}}(p^*)$ and $U_{C_{R}}=U_{\mathscr{S}}(p^*)$ where $U_{\mathscr{R}}(p^*)$ and $U_{\mathscr{S}}(p^*)$ are defined as follows.
          \begin{enumerate}
          \item $U_{\mathscr{R}}(p^*)$ represents the relation between pressure and flow four-velocity behind the \emph{rarefaction} wave:\\
          Given the pressure behind the rarefaction wave (i.e. $p^{*}$) during the Newton-Raphson iteration for solving \Cref{eq:p*}, we can determine $U_{\mathscr{R}}(p^*)$ by numerically solving the system of equations \eqref{TM EOS}, \eqref{sound_speed}, \eqref{eq:p=rhoT}, \eqref{eq:Riemann invariant}, and \eqref{eq:entropy}:
          \begin{subequations}
          \begin{align}
          &\frac{dU}{d\rho}=\pm \frac{c_{\text{s}}\gamma}{\rho},\label{eq:Riemann invariant}\\
          &\frac{p}{\rho^{5/3}}(h-T)=\text{constant}.\label{eq:entropy}
          \end{align}
          \label{eq: U behind r wave}
          \end{subequations}
          Hereafter, the upper/lower sign applies to the right/left wave. The ordinary differential equation \eqref{eq:Riemann invariant}, known as the Riemann invariant \citep{Rezzolla2018}, relates the dynamical ($U$) and thermal ($c_{\text{s}}$) quantities. \Cref{eq:entropy}, derived from \Cref{TM EOS} and the second law of thermodynamics, results from the fact that entropy is constant through the rarefaction wave. The `constant' in \Cref{eq:entropy} is a function of entropy and can be determined by the thermal quantities in the region unperturbed by the rarefaction wave. In the case of the constant polytropic EoS, \Cref{eq:entropy} reduces to a familiar form: $p/\rho^{\Gamma}=\text{const}$.

          \item $U_{\mathscr{S}}(p^*)$ represents the relation between pressure and flow four-velocity behind the \emph{shock} wave:\\
          Let `up/down' denote the upstream/downstream state of the shock wave. Under the condition that $p_{\text{down}}$ ($p_{\text{down}}=p^{*}$ in this case) is given during the Newton-Raphson iteration for solving \Cref{eq:p*}, we can compute $h_{\text{down}}$ by numerically solving the jump condition of the enthalpy:
          \begin{equation}
          h^2_{\text{up}}-h^2_{\text{down}}=\left(\frac{h_{\text{down}}}{\rho_{\text{down}}}+\frac{h_{\text{up}}}{\rho_{\text{up}}}\right)\left(p_{\text{up}}-p_{\text{down}}\right).
          \label{eq:TaubAdiabatic}
          \end{equation}
          \Cref{eq:TaubAdiabatic} is known as the Taub adiabat \citep{Taub}, where $\rho_{\text{down}}$ can be eliminated using Equations \eqref{T of h} and \eqref{eq:p=rhoT}. After determining $h_{\text{down}}$ by a root-finding routine, the mass flux across the shock can be calculated by
          \begin{equation} J=\left(\frac{p_{\text{down}}-p_{\text{up}}}{h_{\text{up}}/\rho_{\text{up}}-h_{\text{down}}/\rho_{\text{down}}}\right)^{0.5}.
          \label{eq:mass flux}
          \end{equation}
          The four-velocities of shock and post-shock then follow from
          \begin{equation}
          U_{\text{shock}}=\pm \left(\frac{J}{\rho_{\text{up}}}\right)\sqrt{1+U^2_{\text{up}}}\pm U_{\text{up}}\sqrt{1+\left(\frac{J}{\rho_{\text{up}}}\right)^2},
          \label{eq:Ushock}
          \end{equation}
          and
          \begin{equation}
          U_{\mathscr{S}}(p^{*})=\mp \left(\frac{J}{\rho_{\text{down}}}\right)\sqrt{1+U^2_{\text{shock}}}+U_{\text{shock}}\sqrt{1+\left(\frac{J}{\rho_{\text{down}}}\right)^2},
          \label{eq:Udown}
          \end{equation}
          respectively. Equations \eqref{eq:Ushock} and \eqref{eq:Udown} are essentially the Lorentz boost that takes four-velocity from the shock rest frame to the lab frame. Note that the mass flux $J$ is an invariant under the Lorentz boost in the flow direction.
          \end{enumerate}

    \item Once $p^{*}$ is known, $\rho_{\text{down}}$ follows from Equations \eqref{T of h} and \eqref{eq:p=rhoT}, which in turn allows for computing $U_{\text{down}}$ and $U_{\text{shock}}$ through \Cref{eq:Ushock} and \Cref{eq:Udown}. On the other hand, $\rho$ behind the rarefaction wave follows from solving the system of equations \eqref{TM EOS}, \eqref{eq:p=rhoT}, and \eqref{eq:entropy}. Finally, given the self-similar and isentropic character of the rarefaction wave, $\rho$ and $U$ within the rarefaction fan can be computed by solving the system of equations \eqref{eq:Riemann invariant} and $U(\xi)=\texttt{Relative}(\xi/\sqrt{1-\xi^2},\pm c_{s}/\sqrt{1-c_{s}^2})$, where $\xi=x/t$.
    \end{enumerate}

Based on the above procedure, we show in \Cref{tb:exact solution} the exact solution of the relativistic Riemann problem given in the last row of Table \ref{tb:IC_RiemannProblems} with the TM EoS at $t=80.0$. The source code is available at (\url{https://github.com/zengbs/ExactSolutionRelativisticRiemannProblem}).

\definecolor{MixedLimitsExact}{RGB}{86, 140, 227}
\begin{table}[]
\caption{
Exact solution of a relativistic Riemann problem with the TM EoS at $t=80.0$. Columns from left to right give $x$-coordinate, proper mass density, four-velocity, and pressure. The initial condition is given in the last row of Table \ref{tb:IC_RiemannProblems}, with the initial discontinuity at $x=5\times10^{-2}$. The blue solid line in \Cref{fig:non-relativistic shock tube} plots the solution. The exact solution is calculated with double precision and shown in 16 digits to reach machine accuracy. The data are available in the supplement.}
\label{tb:exact solution}
\begin{tabular}{@{}cccc@{}}
\toprule
$x$                             & $\rho$                          & $U_{x}$                          & $p$                             \\ \midrule
\texttt{0.0000000000000000e+00} & \texttt{1.0000000000000000e+02} & \texttt{+1.0000000000000000e-03} & \texttt{1.0000000000000000e-04} \\
\texttt{2.5743971630613077e-02} & \texttt{1.0000000000000000e+02} & \texttt{+1.0000000000000000e-03} & \texttt{1.0000000000000000e-04} \\
\texttt{2.6720534130613077e-02} & \texttt{9.9999999999999986e+01} & \texttt{+1.0000000000000002e-03} & \texttt{1.0000000000000000e-04} \\
\texttt{2.8673659130613080e-02} & \texttt{9.8588362795909134e+01} & \texttt{+1.0183105709873300e-03} & \texttt{9.7658360819209613e-05} \\
\texttt{3.1603346630613080e-02} & \texttt{9.6495914226915929e+01} & \texttt{+1.0457764274335814e-03} & \texttt{9.4228343648087098e-05} \\
\texttt{3.6486159130613087e-02} & \texttt{9.3074657510006034e+01} & \texttt{+1.0915528549894106e-03} & \texttt{8.8726314406083176e-05} \\
\texttt{4.2345534130613087e-02} & \texttt{8.9077088828567909e+01} & \texttt{+1.1464845688462534e-03} & \texttt{8.2466343818876140e-05} \\
\texttt{5.2111159130613087e-02} & \texttt{8.2671707042031258e+01} & \texttt{+1.2380374291888151e-03} & \texttt{7.2821829638494934e-05} \\
\texttt{6.4806471630613094e-02} & \texttt{7.4813960019366874e+01} & \texttt{+1.3570561601099081e-03} & \texttt{6.1655409508574801e-05} \\
\texttt{8.4337721630613108e-02} & \texttt{6.3723430244968533e+01} & \texttt{+1.5401619444173906e-03} & \texttt{4.7188055213551995e-05} \\
\texttt{1.1168147163061304e-01} & \texttt{5.0121316453652021e+01} & \texttt{+1.7965101817141935e-03} & \texttt{3.1625521037347636e-05} \\
\texttt{1.5172053413061287e-01} & \texttt{3.3922515604881056e+01} & \texttt{+2.1718776864619303e-03} & \texttt{1.6499866085321606e-05} \\
\texttt{2.0836115913061276e-01} & \texttt{1.7591444669719621e+01} & \texttt{+2.7028867092515813e-03} & \texttt{5.5229310921865207e-06} \\
\texttt{2.0933772163061276e-01} & \texttt{1.7369735883307754e+01} & \texttt{+2.7120420544404751e-03} & \texttt{5.4074080094554571e-06} \\
\texttt{2.0972078270771960e-01} & \texttt{1.7283280852025452e+01} & \texttt{+2.7156332803617649e-03} & \texttt{5.3626249948767070e-06} \\
\texttt{2.6627329885804002e-01} & \texttt{1.7283280852025452e+01} & \texttt{+2.7156332803617649e-03} & \texttt{5.3626249948767070e-06} \\
\texttt{2.6724986145813656e-01} & \texttt{4.0108528993879889e-10} & \texttt{+2.7156332816129858e-03} & \texttt{5.3626249948767070e-06} \\
\texttt{2.6909288248391281e+01} & \texttt{4.0108528993879889e-10} & \texttt{+2.7156332816129858e-03} & \texttt{5.3626249948767070e-06} \\
\texttt{2.6910264810891281e+01} & \texttt{9.9999999999999998e-13} & \texttt{-1.0000000000000000e+02} & \texttt{1.0000000000000000e-10} \\
\texttt{1.0000000000000000e+02} & \texttt{9.9999999999999998e-13} & \texttt{-1.0000000000000000e+02} & \texttt{1.0000000000000000e-10} \\ \bottomrule
\end{tabular}
\end{table}

\end{document}